\documentclass[sigconf,9pt]{acmart}
\usepackage[T1]{fontenc}
\usepackage{hyperref}
\usepackage{ragged2e}
\usepackage{tikz}
\usepackage{booktabs}
\usepackage{amsmath}
\usepackage{filecontents}
\usepackage{adjustbox}
\usepackage{natbib}
\usepackage{xspace}

\usepackage{subcaption}
\usepackage{tikz}
\usepackage{balance}
\usepackage{xcolor}
\usepackage{array}

\usepackage{enumitem}
\usepackage{multirow}
\usepackage{multicol}
\usepackage{color}
\usepackage{soul}
\usepackage{colortbl}
\usepackage{tabularx}
\usepackage{threeparttable}
\usepackage{lscape}

\usepackage{xparse}
\NewDocumentCommand{\Log}{o}{%
  \IfNoValueTF{#1}{}{{}^{#1}\!}\log}%
  
\usepackage{balance}
\usepackage{url}

\usepackage{breakurl}

\newcommand{\todo}[1]{}
\renewcommand{\todo}[1]{{\color{orange} TODO: {#1}}}

\newcommand{\todoKath}[1]{}
\renewcommand{\todoKath}[1]{{\color{blue} Kath:  {#1}}}

\newcommand{\Kimb}[1]{}
\renewcommand{\Kimb}[1]{{\color{green} KIMB:  {#1}}}

\newcommand{\system}{GPS\xspace}

\newcommand{\probP}{\text{I\kern-0.15em P}}

\newcommand{\syn}{\textls[-30]{\textsc{syn}}\xspace}
\newcommand{\JOIN}{\textls[-30]{\textsc{JOIN}}\xspace}

\date{}

\title{Predicting IPv4 Services Across All Ports}
\author[]{Liz Izhikevich}
\affiliation[]{Stanford University}

\author[]{Renata Teixeira}
\affiliation[]{Inria, Paris}

\author[]{Zakir Durumeric}
\affiliation[]{Stanford University}

\copyrightyear{2022} 
\acmYear{2022} 
\setcopyright{acmlicensed}\acmConference[SIGCOMM '22]{ACM SIGCOMM 2022 Conference}{August 22--26, 2022}{Amsterdam, Netherlands}
\acmBooktitle{ACM SIGCOMM 2022 Conference (SIGCOMM '22), August 22--26, 2022, Amsterdam, Netherlands}
\acmPrice{15.00}
\acmDOI{10.1145/3544216.3544249}
\acmISBN{978-1-4503-9420-8/22/08}


\begin{document}

\begin{abstract}
Internet-wide scanning is commonly used to understand the topology and security of the Internet. However, IPv4 Internet scans have been limited to scanning only a subset of services---exhaustively scanning all IPv4 services is too costly and no existing bandwidth-saving frameworks are designed to scan IPv4 addresses across all ports. 
In this work we introduce \system, a system that efficiently discovers Internet services across all ports.
\system runs a predictive framework that learns from extremely small sample sizes and is highly parallelizable, allowing it to quickly find patterns between services across all 65K ports and a myriad of features.
\system computes service predictions in 13~minutes (four orders of magnitude faster than prior work) and finds 92.5\% of services across \textit{all} ports with $131\times$ less bandwidth, and $204\times$ more precision, compared to exhaustive scanning.
\system is the first work to show that, given at least two responsive IP addresses on a port to train from, predicting the majority of services across all ports is possible and practical.

\end{abstract}

\maketitle
\section{Introduction}

Internet-wide scanning allows researchers and network operators to understand how the Internet works \textit{in practice}, and has been used to study network topology~\cite{vermeulen2020diamond,beverly2016yarrp}, operator decisions~\cite{li2016you,durumeric2014matter} and security vulnerabilities~\cite{beurdouche2015messy,checkoway2014practical}.
Internet-wide scanning works by initiating network connections with services on a set of given ports. 
Unfortunately, no study has been able to analyze the entire IPv4 service space across all ports, as scanning all 65K ports across all 3.7~billion IPv4 addresses would require 5.6~years using ZMap~\cite{durumeric2013zmap} at 1~Gbps---a scanning rate that prevents flooding destination networks. 
As a result, Internet-wide studies often scan only a relevant subset of assigned ports (e.g., 23/TELNET and 2323/TELNET), and popular Internet-service ``search engines'' like Censys~\cite{durumeric2015search} and Shodan~\cite{shodan} resort to scanning only the most populated ports. 

However, recent work has shown that the majority of Internet services do not run on assigned ports. Scanning only port~80 misses 97\% (1.8~billion) of all IPv4 HTTP services and scanning only port~23 misses 95\% (6.8~million) of Telnet services~\cite{lzr}.
Services occupy a long tail of nonstandard ports~\cite{lzr}:
Internet service search engines 
that scan the five thousand most populated ports (e.g., Censys)
miss the majority---an estimated 1.9~\textit{billion} (63\%)---of Internet services. 
To complicate matters, the services on non-standard ports are not accurately represented by those on standardized ports: IoT and security-critical devices are up to five times more likely to inhabit the 1.9~billion services on non-standard ports~\cite{lzr}. 
Unfortunately, botnets already target vulnerable services on non-standard ports~\cite{mirai}.
Further, 
both researchers and network operators often cannot rely on sub-sampling when needing to find a ``needle in the haystack'' in the long tail of Internet services.
For example, Marczak et~al.\ rely on complete Internet-wide scans of select ports to identify spyware infrastructure~\cite{marczak2021candiru} and clients of cyber espionage~\cite{marczak2020running}, which occupy only a few hundred compromised servers.
For researchers and operators to secure the entire Internet, it is imperative to find services across {all} IP addresses \textit{and} ports. 

In this work, we introduce \system, an intelligent scanning system that scalably and efficiently finds services across all IPv4 addresses and ports.  \system assumes no prior knowledge and starts by collecting an initial set of ``seed'' data. It uses a myriad of application, transport, and network layer features to probabilistically model service presence. \system uses the model to construct a two-phased scanning approach that: (1) finds at least one service on all hosts, and (2) uses the discovered service to find remaining services on the same host. \system's key contribution is a parallelizable and accurate predictive algorithm---relying on calculating simple conditional probabilities between features and services---that can find patterns and predict new services across all ports in parallel with minimal training data. 


\system finds 92\% of services across all 65K ports (with greater than two responsive IP addresses) using 131~times less bandwidth and 204~times higher precision than exhaustive scanning. We show that as services become harder to predict, the bandwidth required to find new services increases: finding 96\% of services across all ports uses only 10~times less bandwidth compared to exhaustively scanning.  
When scanning only popular ports, \system's algorithm uses up to 28~times less bandwidth than the state-of-the-art ML-based solution~\cite{sarabi}.  
On a single CPU core, \system computes predictions five times faster than if prior work was extended to predict services across all ports.
When using elastic cloud platforms, \system's parallelizeable algorithm computes predictions in 13~minutes---four orders of magnitude faster than extending prior work.

\system is the first practical system that finds the majority of services across all ports on the IPv4 address space.
We are releasing \system under the Apache 2.0 License so that individual researchers, network operators, and security companies can reduce the needed bandwidth to find, study, and secure the majority of exposed---and perhaps already exploited---services on the Internet. 
\section{Prior Work}
\label{sub:sec:prior}

To reduce the cost of scanning and identify additional services, prior work has proposed several solutions to predict which IP addresses to scan on a given port.
There exist two drawbacks across prior work: (1) no system is designed---nor scales---to predict services across \textit{all} 65K~ports, and (2) existing solutions require a prohibitive amount of training data, which would take years to collect due to the sparse search space.

\vspace{3pt}
\noindent
\textbf{Classifiers.}\quad
Sarabi~et~al.\ ~\cite{sarabi} approach intelligent Internet-wide scanning as a classification problem---in which each port value is a class---and use an XGBoost classifier~\cite{chen2016xgboost} to predict whether an IP will respond on one of 20~popular ports. 
Their work finds that the strongest predictor of a service is the presence of other services on the same host. To find patterns across common ports, their system sequentially trains a model per port and uses the output of each model to train the next.
Unfortunately, their method does not scale to all 65K~ports for two reasons.
First, their model requires at least 10~million services to train on per port, which is not available across 99.99\% of ports (many ports simply do not have that number of services present). 
Second, an individual model must be trained for each port, making the computation time prohibitively expensive: the training and prediction process per port requires roughly 70~seconds on an NVIDIA GeForce GTX 1080 Ti GPU, and requires roughly 53~days of computation across all 65K~ports. This computation cannot be parallelized, because the training of the models is sequential. We compare the performance of XGBoost scanner to \system in Section~\ref{sub:sub:sec:sysML}, and show that \system is more accurate than XGBoost on average, is capable of predicting services across all ports, and takes four orders of magnitude less time to compute service predictions. 

\vspace{3pt}
\noindent
\textbf{Target generation algorithms.}\quad
Previous work~\cite{murdock2017target,foremski2016entropy,gasser2018clusters} has predicted services on IPv6 addresses using \textit{Target Generation Algorithms} (TGAs). TGAs learn the structure of known IP addresses and predict similarly structured addresses that are likely to run services. However, TGAs are unable to find patterns across all 65K ports and require a new model to be built and trained for each port.  
TGAs also rely solely on subnetwork correlations, which we show are orders of magnitude less predictive of hosts on uncommon ports (Section~\ref{sub:predictability}).
Most importantly, TGAs are not computationally scalable: obtaining the minimum number of IP addresses required to effectively train the model (i.e, 1,000~IPs~\cite{foremski2016entropy}) across 90\% of ports would require randomly probing at least 25\% of the address space per port---due to the sparsity of most uncommon services---requiring over one year to collect using ZMap~\cite{durumeric2013zmap} at~1Gbps.

We verify whether TGAs are capable of predicting services on IPv4 addresses across densely-populated ports by modifying Entropy/IP~\cite{foremski2016entropy} and EIP~\cite{gasser2018clusters} to predict IPv4 addresses (predicting one IPv4 octet at a time instead of one IPv6 nibble). We use 1,000 randomly sub-sampled addresses from Censys' Universal Internet Dataset containing 100\% IPv4 scans across 2K ports to train a model for each port. Each model predicts 1M candidate addresses per port (an order of magnitude more addresses than the number of IPs that respond across 90\% of ports). Combining all candidate addresses, Entropy/IP and EIP are able to find only 19\% of services in the Censys dataset. 

\vspace{3pt}
\noindent
\textbf{Recommendation Systems.}\quad
Proprietary recommendation systems have been successful at recommending millions of items to millions of users~\cite{gomez2015netflix,mcinerney2018explore,ma2020temporal}. We apply a popular open-sourced hybrid recommender~\cite{kula2015metadata} to recommend candidate ports to IP addresses and find it to be unsuccessful at predicting services (Appendix~\ref{app:rec_sys}).

\section{\system Objective}
\label{sub:sec:sys_objective}


\system's objective is to maximize finding services across \textit{all} ports. 
This differs from prior Internet scanning systems~\cite{sarabi} that maximize the total ``fraction of services'' found (Equation~\ref{eq:frac_services}): the number of services found by the system relative to the number of IPs that are found in a ``ground truth'' set (i.e., a baseline for the true number of services on port $p$, obtained by a 100\% scan). 
This metric is biased towards discovering services on popular ports where services are more dense (e.g., 5\% of all services across all 65K ports live on the top~10 ports) and disincentives finding services on uncommon ports, which are understudied and more likely to be vulnerable~\cite{lzr}. 

\begin{equation} \label{eq:frac_services}
\mbox{ Fraction of Services = }  \frac{\mbox{${\# (IP, p)}$ Found by System}}{\mbox{${\# (IP, p)}$ in Ground Truth}}
\end{equation}

\vspace{2pt}
\noindent
To address this, we introduce an additional \textit{normalized service} metric (Equation~\ref{eq:norm_services}) which, given port $p$, normalizes the weight of a service based on the number of IPs that respond on port $p$ (i.e., $IP_p$) in a ``ground truth'' set. 
By doing so, discovering all services on an uncommon port holds equal weight compared to discovering all services on a popular port. 
Using the new metric, \system's goal (Equation~\ref{eq:opt_form}) is to maximize the fraction of normalized services found across the set of all ports ($|\mathcal{P}|$), while constraining the number of probes (i.e., bandwidth) by constant $c_1$.  

\begin{equation} \label{eq:norm_services}
\mbox{ Normalized Services = }  \frac{\sum_{p \in \mathcal{P}}\frac{\mbox{${\# IP}_{p}$ Found by System}}{\mbox{${\# IP}_{p}$ in Ground Truth}} }{|\mathcal{P}|}
\end{equation}

\begin{equation} \label{eq:opt_form}
\begin{aligned} 
\max \, & \mbox{Normalized Services}(bandwidth) \,\,\,\,\,\,\,\,\,\,\,\,\,\,\,\,\,\,\,\,\,\,\,\,\,\,\,\,\,\,\,\,\,\,\,  \\
 & bandwidth < c_1 
 \end{aligned}
\end{equation}

\noindent By constraining bandwidth, \system is optimized to scan only the most predictable services in order to maximize the total fraction of normalized services found. The more bandwidth available, the more services \system finds. 

There are two additional constraints when building a deployable system:
\vspace{2pt} \\
(1) \textbf{Computationally scalable:} Predicting services across all ports and IP addresses encompasses a large search space, but requires a solution that operates in a constrained amount of wall-time. 
This is especially important as Internet services are constantly changing~\cite{moura2015dynamic}. For example,
we conduct a scan of the same 0.1\% of the IPv4 address space across 65K~ports on June 17, 2021 and June 27, 2021. Within the 10-day period, 15\% of normalized services and 9\% of all services 
disappear. Thus, if predicting services takes too long, the initial set of services that the model uses to learn patterns from, and the model's predictions, will likely become obsolete. 
\vspace{2pt}
 \\ 
(2) \textbf{No prior knowledge:} Intelligent Internet scanning is not a classic prediction task in which informative features are initially available for training. 
Predictive features reside on IP addresses and ports that are not initially known. 
Since \system starts with no prior information about the location of Internet services, it must
devise an efficient scanning strategy that gradually discovers information about hosts, which can later be used to predict other services. 

\vspace{5pt}
\noindent
\textbf{Ethics.}\quad
Any open source scanning technique can be used both by researchers to monitor and secure Internet services, and by attackers to uncover vulnerabilities. We design \system to be easily blocked by network operators (Section~\ref{sub:sec:implementation}), which disincentives attackers from using \system while enabling researchers to gain visibility of and secure the Internet.
When executing scans and building \system, we follow the community standards for good Internet citizenship outlined by Durumeric et~al.~\cite{durumeric2013zmap}. We also emphasize that \system reduces the traffic sent when scanning services on a large number of ports (Section~\ref{sec:alg_eval}), thereby reducing the impact on destination networks. We hope that researchers and companies both consider using \system to reduce the total number of scan probes on the Internet.

\section{Identifying Predictive Features} 
\label{sub:predictability}
\label{sub:sub:sec:net_patterns}
\label{sub:sec:features}



At first glance, finding IPv4 services  across 3.7~billion addresses and 65K~ports---a nearly $2^{48}$ search space---may seem intractable. 
However, prior studies have surfaced several ways to predict the locations of \textit{popular} services. We begin by investigating whether these predictive patterns can be used to predict services across all 65K~ports.  
Our results build the foundational set of three categories of predictive features that \system will rely on: network layer, transport layer, and application layer. 

\vspace{3pt}
\noindent
\textbf{Port usage is correlated (Transport layer).} \quad
Bano et~al.\ showed that among the eight most popular ports, the presence of one port on a host can be used to predict the presence of other ports~\cite{bano2018scanning}.
We scan all 65K~ports on a 1\% random IPv4 sample in March 2021 and find the same holds true across the majority of all ports: for every port, at least 25\% of hosts also respond on the same second port. 
Thus, a service's port can be used to predict other open ports on the same host. 

\vspace{3pt}
\noindent
\textbf{Different populations of hosts are more likely to run specific services (Application layer).}\quad
Prior work~\cite{kumarThings,lzr} shows that IoT and router vendors often manufacture particular ports to be open in order to provide network access. Thus, application-layer data (e.g., TLS certificates) that indicates manufacturer or operating system can likely be used to predict other services on the same host. 
Using the same scan, we manually investigate the top 3K most common HTTP header values, SSH banners, and TLS certificates across all hosts. 
We find that IoT devices and routers are the most popular host type across the majority of ports. Having devices with manufactured---and thus predictable---port presence dominate services across all ports encourages the pursuit of predicting services on uncommon ports. 
Application layer features that identify a host's manufacturer (e.g., the organization, subject name or issuer of a TLS certificate, SSH Banner, PPTP vendor, etc.), operating system (e.g., HTTP Server, SSH banner, CWMP Header, MySQL server version, etc), purpose (e.g., HTTP HTML title, VNC desktop name, etc), and owner (e.g., SSH key, TLS certificate, etc.) can predict service presence. 

\vspace{3pt}
\noindent
\textbf{Internet services are more likely to appear together in networks (Network layer).}\quad
Murdock et~al.\ and Foremski et~al.\ found that hosts in a given network are more likely to have the same ports open~\cite{murdock2017target,foremski2016entropy}. 
Using Censys' universal data set comprising of 100\% IPv4 scans of over 2K ports~\cite{censysUniversal}, we find that 81\% of all services appear at least twice on the same port within the same /16 subnetwork. This result indicates that the network is predictive of service presence. However, as ports become less popular, the probability of finding two services responding on the same port in the same /16 subnetwork becomes as low as 0.02\%. 
Thus, while network patterns are effective at finding hosts with a popular port open, discovering services on uncommon ports cannot be done by solely relying on network patterns.
\vspace{5pt}

\noindent In Section~\ref{sub:sub:sec: identify_important}, \system uses transport-layer, application-layer, and network-layer features to predict service presence.

\section{System Architecture}

In this section, we introduce \system, a system that efficiently finds Internet services across all ports. 
\system assumes no prior knowledge and uses a four phase process to bootstrap itself and comprehensively discover services. 
First,  through random sampled scanning, \system collects a ``seed set'' of Internet hosts and services to learn from. 
Second, \system creates a probabilistic model of the most predictive feature values across all services in the seed set. 
Third, because \system's seed set spans only a subset of hosts, \system uses its probabilistic model to find at least one service on each responsive IPv4 host. 
Finding the first service of a host is non-trivial and bandwidth expensive, but is crucial to predicting remaining services.
Fourth, once \system has discovered at least one service on each responsive host, \system predicts remaining services.

\system's key insight is using a simple and parallelizable computation for predicting services that is not dependent upon a large training set.
While \system's algorithm is computationally expensive, we provide an implementation that marries serverless computing's elastic resources with \system's parallelization, speeding up \system's wall-clock prediction time by orders of magnitude.




\subsection{Building a Seed Set}
\label{sub:sec:filter}
\system starts with no knowledge about Internet hosts. To learn patterns and find services, \system must either collect, or use an available (e.g., the LZR dataset~\cite{lzrData}) uniform random sample of IPv4 hosts each scanned across all 65K ports (i.e., ``a seed set'').  
The size of the seed scan directly impacts prediction quality and is the primary way to ensure that \system will learn all predictive patterns.
We show in Appendix~\ref{app:seedsize} the bandwidth/coverage trade-off when determining the size of a seed-scan: larger seed sizes find more services, but use substantially more bandwidth. 
Importantly, unlike prior work (Section~\ref{sub:sec:prior}) that requires large sample sizes (e.g., 25\% IPv4 sample) to train, \system is able to predict services across the majority of ports with only a 0.1\% IPv4 sample.
\system allows the user to specify the desired size of the seed scan as an input parameter based on their bandwidth constraint (Equation~\ref{eq:opt_form}).


\begin{table}[t]
\centering
\small
\begin{tabular}{ll}
\toprule
Application-Layer or & \# Unique Values in  \\ 
Network-Layer Feature & Censys Ground Truth  \\ 
\midrule

Protocol & 56 \\
TLS Cert: Hash & 30.1M \\
TLS Cert: Organization & 1.1M\\
TLS Cert: Subject Name & 27.9M \\
HTTP: HTML title& 5.9M \\
HTTP: Body Hash& 50.8M \\
HTTP: Server & 480K\\
HTTP: Header & 22K\\
SSH: Host Key & 14.3M\\
SSH: Banner & 177K\\
VNC: Desktop Name& 4.5K \\
SMTP: Banner & 2.9M \\
FTP: Banner & 1.5M\\
IMAP: Banner & 144K\\
POP3: Banner& 390K \\
CWMP: Header& 10 \\
CWMP: Body Hash& 11\\
Telnet: Banner& 219K \\
PPTP: Vendor& 390K \\
MYSQL: Server Version & 5.7K\\
Memcached: Server Version& 129 \\
MSSQL: Server Version & 381\\
IPMI: Banner& 116 \\
\midrule
IP's /16 subnetwork & 37.3K \\
IP's ASN & 67.7K \\

\bottomrule
\end{tabular}
\vspace{3pt}
	\caption{\textbf{ \system Features}---%
\textnormal{
\system's features span all TCP protocols with an available banner on Censys (i.e., 15~unique protocols). \system only uses network features that are most predictive of service presence; the filtering process is described in Appendix~\ref{app:features}. The dimensionality (i.e., number of unique values) is calculated using the Censys ground truth dataset described in Section~\ref{sub:sec:exp}.}
	 }
	 \label{table:predictive_features_app}
\end{table}

\subsection{Identifying Predictive Patterns}
\label{sec:sub:scanning_algorithm}
\label{sub:sub:sec: identify_important}

\system's second phase is purely computational: building a probabilistic model to identify feature patterns most predictive of service presence. \system uses the probabilistic model in the third and fourth stages of the process to scan for new services (Sections~\ref{sub:sec:optimal_scanning} and~\ref{sub:sec:predicting}). 

\vspace{3pt}
\noindent
\textbf{Feature selection.}\quad
To identify feature values that are most predictive of service presence, \system uses the seed set to extract the three categories of features---application, transport, and network---introduced in Section~\ref{sub:predictability}.
\system uses a total of 25~unique application and network layer features (Table~\ref{table:predictive_features_app}) in our evaluation. 
\system's features span all TCP protocols with an available banner on Censys (i.e., 15~unique protocols). \system therefore accommodates the possibility that both popular (e.g., HTTP) and less popular (e.g., VNC) protocols contain information that are predictive of service presence. 
\system's design allows for the user to easily include/remove any feature candidates. 

\vspace{3pt}
\noindent
\textbf{Modeling interactions.} \quad
\system independently models different \textit{interactions} of the three primary feature categories. This accommodates the possibility that any combination of the three feature categories may be predictive of Internet services. 
For example, knowing a host's network ($Net_{IP}$) can be less predictive when the host appears in many subnetworks (e.g., Android TVs) or more predictive when a service is in only one subnetwork (e.g., Freebox devices only appear in the Free network~\cite{freebox}).
Concretely, \system models the following:

\vspace{5pt}
\noindent
1. Transport layer correspondence: the probability that a host will have $Port_{a}$ open given the host has $Port_{b}$ open (Expression~\ref{eq:l4}).
\begin{gather}
\probP( Port_{a} | Port_{b} )  \label{eq:l4}
\end{gather}
2. Transport and application layer correspondence: the probability that a host will have $Port_{a}$ open given the host has $Port_{b}$ open and the service on $Port_{b}$ contains a specific application layer feature value  (Expression~\ref{eq:l4_7}).
\begin{gather}
\probP( Port_{a} | ( Port_{b}, App_{Port_{b}})  )  \label{eq:l4_7} 
\end{gather}
3. Transport and network correspondence: the probability that a host will have $Port_{a}$ open given the host has $Port_{b}$ open and the host responding on $Port_{b}$ resides in network $Net_{IP}$ (Expression~\ref{eq:l4_host}).
\begin{gather}
\probP( Port_{a} | ( Port_{b}, Net_{IP})  ) \label{eq:l4_host} 
\end{gather}
4. Transport, application and network correspondence: the probability that a host will have $Port_{a}$ open given the host has $Port_{b}$ open and the service on $Port_{b}$ contains a specific application layer feature value and the host responding on $Port_{b}$ resides in network $Net_{IP}$ (Expression~\ref{eq:l4_l7_host}).
\begin{gather}
\probP( Port_{a} |( Port_{b}, App_{Port_{b}},Net_{IP})  ) \label{eq:l4_l7_host}
\end{gather}

\noindent
Note that all conditional probabilities rely on $Port_{b}$ having already been discovered, such that the relevant application layer and/or transport layer feature values can be used to predict $Port_{a}$. 
However, not all hosts respond on more than one port and not all hosts have a priori available information (e.g., a known service). We show in the following section how \system uses its probabilistic models to predict services no matter how many ports the hosts responds on or how much a priori host information is available. 

\vspace{3pt}
\noindent
\textbf{Computational scalability.}\quad
We note that \system's method for modeling feature interactions is computationally expensive; \system must calculate all possible combinations of features that appear in the seed set. Furthermore, while the port feature ($Port_{a}$) has a dimensionality of 65K, application-layer features can vary widely in dimensionality (Table~\ref{table:predictive_features_app}), thus allowing for potentially billions of feature-value combinations. 
Nevertheless, \system's method is computationally scalable across all 65K ports because computing conditional probabilities is a parallelizable computation across all 65K ports. 
We discuss our implementation of the probabilistic model in Section~\ref{sub:sec:implementation} and how we use serverless computing to dramatically reduce the wall-clock time and use matrix operations to scale computations. 
We formally evaluate the computational complexity of \system in Section~\ref{sub:sec:comp_compl}.

\subsection{Predicting The First Service}
\label{sub:sub:prior_know}
\label{sub:sec:optimal_scanning}

Predicting a host's first service must be treated differently than predicting the remaining services on the host.
When predicting a host's first service, only network-layer features (i.e., the IP addresses' subnetwork) are available to predict service presence. In contrast, when at least one service is known, the information provided by that service can be used to predict other services on the same host (e.g., port correlations).  
Since the seed set only contains information about a subset (e.g., 1\%) of all hosts, \system must first discover at least one service on each responsive IPv4 host in order to use it to further predict other services on the same host.


\vspace{3pt}
\noindent
\textbf{Scanning step size.}\quad
As presented in Section~\ref{sub:predictability}, services are more likely to appear together in subnetworks. Thus, given a responsive service $(IP,p)$ in the seed set, \system must probe the IP addresses around $IP$ (i.e., in its network) on port $p$, to maximize its chances of finding other services. 

\system faces a trade-off when building an efficient scanning strategy: how exhaustively should the subnetwork of a responsive service in the seed set be scanned? 
Scanning 100\% of the IPv4 address space on a port increases the likelihood of finding all hosts that respond on that port, but requires more bandwidth and increases the impact on destination networks.
Scanning, for example, the /24 subnetwork of IP $IP$ on port $p$ from the seed set requires substantially less bandwidth, but potentially misses hosts outside of the subnetwork that also respond on port $p$.
The user's bandwidth constraint is the deciding factor of how large of a ``scanning step size'' (i.e., the subnetwork size to exhaustively scan a port) \system should use, and is left as a user-specified parameter. 
We formally evaluate the bandwidth/coverage trade-off in Section~\ref{sub:sec:param} to help inform the user what scanning step size to use. 

Relying on selective random probing of a particular network and port is a bandwidth-expensive but initially unavoidable process. Thus, \system uses it only when absolutely necessary: predicting on each host \textit{only} the service(s) that must be found first in order to predict any remaining services.

\vspace{3pt}
\noindent
\textbf{Choosing the most predictive services.}\quad
\system prioritizes finding the minimum set of services that is most informative for predicting any remaining services. 
For example, for a host that responds with a generic HTTP page on port~80 and a vendor-revealing banner on port~222, discovering the banner is likely much more predictive of the HTTP/80 service than vice versa (most HTTP/80 services do not respond on SSH/222 but most SSH/222 services do respond on HTTP/80). 
\system uses its probablistic models (Equations~\ref{eq:l4}--\ref{eq:l4_l7_host}) to calculate which service for each host in the seed set is most predictive of all the host's remaining services. If a host only responds on one port in the seed set,
the sole service is the first and only service that must be predicted.

\system executes the following algorithm to determine which subnetworks and ports to scan in order to prioritize finding the minimum set of most predictive services:
\begin{enumerate}
    \item For all hosts that respond on only one port in the seed set, save the service's $( Port_{a}, Net_{IP})$. 
    \item For all hosts that respond on more than one port in the seed set, compute all four conditional probabilities (Equations~\ref{eq:l4}--\ref{eq:l4_l7_host}). For every $(IP,Port_{a})$ in the seed set, identify the $Port_{b}$ that results in the maximum  \probP($Port_a$), and save the $( Port_{b}, Net_{IP})$.
    \item Across all hosts, group together the required $( Port, Net_{IP})$ tuples and count the number of unique services they help predict in the seed scan (i.e., maximal coverage). 
    \item Sort the $( Port, Net_{IP})$ based on maximum coverage. 
\end{enumerate}

\vspace{3pt}
\noindent
\textbf{Priors scan list.}\quad
\system's algorithm outputs a ``priors scan list'': an ordered list of unique tuples, consisting of a (port, subnetwork of size scanning step) pair (e.g., (80, 1.1.0.0/16)) in order of maximum coverage.
The prior scans list allows \system to collect the most predictive services across all ports in parallel.
Note that not all ports and subnetworks will be scanned: only the (port, subnetwork) tuples that result in the maximum probability of all remaining services being found.
Scanning the priors list allows \system to find the most predictive service on each host, which it can next use to predict additional services on each host.



\subsection{Predicting Additional Services}
\label{sub:sec:predicting}

After \system finds at least one service per host, \system extracts three categories of features---application, transport, and network---in each discovered service to predict additional new services on the same host. 
\system predicts new services by using the existing probabilistic models (Equations~\ref{eq:l4}--\ref{eq:l4_l7_host}) to create a ``most predictive features'' list, and uses that list to predict additional services. 

\system uses the following algorithm to predict additional services:
\begin{enumerate}
    \item For each service in the seed set, identify the feature-tuple (e.g., ($Port_b$, $ App_{Port_{b}}$)) that results in the maximum \probP($Port_a$).  Save the feature tuple and predicted port, $Port_a$, in a ``most predictive feature values'' list. To account for services that appear on random ports and have a low maximum  \probP($Port_a$)), discard all probabilities below 0.00001, which is roughly the hit rate of randomly probing the majority of ports.
    \item For each responsive service found in the \textit{priors} scan, extract available feature values (e.g., TLS certificate, SSH host key).
    \item For each responsive service in the \textit{priors} scan,  and for all of its feature values, if the feature appears in the ``most predictive feature values'' list (i.e., $Port_b$, $App_{Port_{b}}$), save the predicted port  $Port_a$ and the host's IP to a ``predictions'' list.
\end{enumerate}

\noindent
\system's algorithm outputs a ``predictions list'': an ordered list of unique IP addresses and ports to scan. 
Step~1 is crucial to the \system's algorithm; by using every service in the seed scan to build the ``most predictive features'' list,
\system's prediction algorithm guarantees that every service---that \system has seen before in the seed set and is ``predictable''---will be predicted by using the feature value pattern most likely to find the service. 


\subsection{Implementation}
\label{sub:sec:implementation}

We describe an implementation of \system that capitalizes on its parallelizable algorithm to drastically reduce the wall-clock time of predicting services. 

To perform Internet-wide scans (e.g., collect a seed scan, scan for prior services, scan for predicted services), \system chains together three existing tools to conduct a scan: ZMap~\cite{durumeric2013zmap} + LZR~\cite{lzr} + ZGrab~\cite{zgrab2}.
ZMap is a stateless Layer~4 scanner (i.e., ``\syn-scanner'') that initiates TCP connections with Internet hosts in the \system pipeline.
\system's use of ZMap allows it to easily be blocked by network operators, due to its unique fingerprint (IP ID = 54321). 
LZR then takes over the TCP connection, filters out middleboxes, and efficiently fingerprints services (a necessary step when scanning unassigned ports).
For all services that are fingerprinted to be running real services, \system provides LZR the option to forward the connection information to ZGrab, which can then complete the full Layer~7 handshake to collect additional application layer features. 

Implementing \system's algorithm for identifying predictive patterns (Section~\ref{sec:sub:scanning_algorithm}) presents a challenge: finding all pairwise combinations of features and ports---although parallelizable---is computationally and memory intensive.
Thus, to minimize wall-clock time, \system can directly benefit from
a highly parallelizable execution environment that, ideally, would be available to any user of the system. 
While \system can be deployed on any server infrastructure, we find that  
Google BigQuery, a serverless database service that enables scalable analysis of petabytes of data~\cite{bigQuery}, is well suited for the task.
Implementing \system's algorithms  (e.g., calculating conditional probabilities, identifying the most predictive features) in a database query language is a natural choice, as the algorithms rely heavily on reading data, aggregating, and joining among shared data fields.
We formally evaluate the implementation in Section~\ref{sub:section:eval_system_implenetation} and show how with and without a highly parallelizable environment, \system is 5870$\times$ and 5$\times$ faster than prior work, respectively. 

\begin{figure}[t]
  \includegraphics[width=0.5\textwidth]{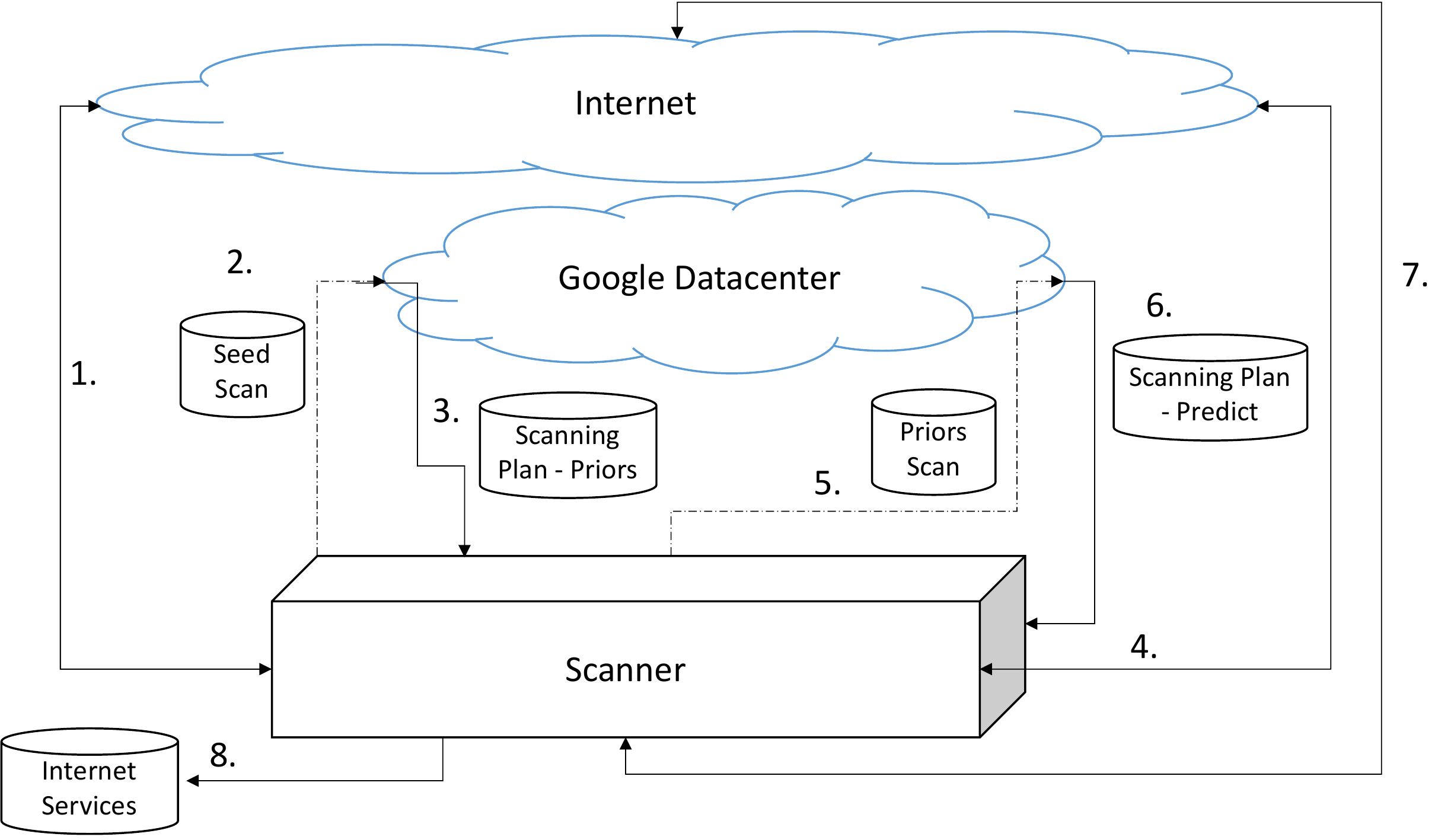}
 \caption{\textbf{\system Implementation}---%
 \textnormal{\system's parallelizable prediction algorithm takes advantage of Google BigQuery's serverless platform to compute predictions in a highly parallelizable execution environment. }}
  \label{fig:system_interaction}
\end{figure}


We illustrate \system's interaction with BigQuery in Figure~\ref{fig:system_interaction}; \system uses BigQuery as a computational engine and does not rely on any pre-existing BigQuery data. 
Upon the completion of the seed scan, \system uploads the results of the scan to BigQuery.
Service features are extracted in BigQuery by either (1)~selecting the appropriate fields from the uploaded scans, (2)~performing operations on the IP address to extract the host's subnetwork, or (3)~ joining on a database that provides the feature (e.g., ASN). 
To efficiently calculate conditional probabilities, \system uses BigQuery's SQL language to compute the pairwise co-occurence matrix for every feature and port, which involves \JOIN-ing the dataset on itself to find all pair-wise combinations of features and aggregating together identical feature value patterns and target-ports to calculate the conditional probabilities. 
\system also uses BigQuery's SQL language to implement the strategy for predicting the first service (Section~\ref{sub:sec:optimal_scanning}), by joining the seed scan with a computed co-occurence matrix. 

Once the strategy for predicting the first service is calculated, \system downloads the strategy on its scanning host. The scanning host exhaustively scans the (port, subnetwork) tuples and uploads the results of the scan to BigQuery, which again extracts the host features. 
\system uses BigQuery's SQL language to create the most predictive feature list (Section~\ref{sub:sec:predicting}) by joining the priors scan with the computed co-occurence matrix, and downloading the prediction strategy (i.e., a list of IPs and ports)  for all remaining services to the scanning host. The scanning host then scans all remaining predicted services, and saves the results in a local database. 
We have released \system's implementation under the Apache~2.0 License at \url{https://github.com/stanford-esrg/gps}.

\section{Evaluation}
\label{sec:alg_eval}

In this section, we evaluate 
\system's ability to find services across all ports. 
First, we evaluate \system's ability to find the maximum number of services under a variable bandwidth constraint. 
We show that, as services become harder to predict, the bandwidth required to find services increases:
\system finds 92\% of services across all ports (with greater than two responsive IP addresses) with $131\times$ less bandwidth compared to exhaustive scanning, but saves only $10\times$ the bandwidth when finding 96\% of services across all ports. 
Second, we evaluate \system's precision. When minimizing Internet scanning's impact on available Internet resources, we show \system is two orders of magnitude more precise than exhaustive probing. 
Third, when scanning popular ports, we compare \system's accuracy and bandwidth with the XGBoost scanner~\cite{sarabi}. \system saves up to 28~times, and on average 2.3~times, the bandwidth required to achieve the same coverage of services. 
Fourth, we evaluate the computational complexity of predicting services and show that, compared to XGBoost scanner, \system computes predictions 5~times faster when using a single core and up to \textit{four orders of magnitude} faster when using BigQuery.
Lastly, using \system's predictions, we identify the feature values that are most predictive of service presence.


\subsection{Methodology}
\label{sub:sec:exp}

Properly evaluating \system requires finding an appropriate ground truth dataset and appropriately configuring \system. 


\vspace{3pt}
\noindent
\textbf{Approximating ground truth.} \quad
Finding a ``ground truth'' dataset to evaluate \system presents a challenge: no method exists to efficiently scan all 65K ports at 100\% (hence the need for \system). Thus, we
evaluate \system using two data sets.
First, we use Censys' universal data set comprising of 100\% IPv4 scans of the most popular 2K ports~\cite{censysUniversal} sampled on July 26, 2021, which is the largest publicly available dataset that scans the most number of ports at 100\%. 
Censys shares that, given their scanning bandwidth, it would require 196~days to exhaustively scan all 65K~ports.

Censys targets only the most popular ports. 
Thus, we additionally evaluate \system against a 1\% random scan of the IPv4 address space across all 65K ports in April, 2021, requiring all 30~days to collect, using the LZR scanner~\cite{lzr}. While the LZR dataset spans all ports, it reduces the available sample for each port, which may miss patterns exhibited by a small number of hosts. 
We filter both dataset for real services using the steps described in Appendix~\ref{app:filt_services}.

To create seed-scans and test sets for each dataset, we randomly assign each IP address, and its accompanying services, to either a seed or test set. 
Thus, a 2\% Censys seed set leaves a 98\% test set, and a 0.5\% LZR seed set leaves a 0.5\% test set. 
Although we spend an entire month scanning 1\% of the IPv4 address space across 65K ports, responsive services are still sparse across most ports. Since we do not expect \system to learn and predict services on ports with no training data, we filter both the LZR seed and test set to only include ports that have greater than two responsive IP addresses. When using a 0.5\% seed set, filtering leaves a remaining 13,162~ports.

\vspace{3pt}
\noindent
\textbf{Parameter tuning.} \quad
GPS requires specifying a seed and step size as input parameters. We evaluate how the seed and scanning step size impact \system's performance and find that a small step size increases \system's precision, but decreases recall (Appendix~\ref{app:parameter_tune}).
This trade off happens because, as discussed in Section~\ref{sub:sub:prior_know}, scanning a smaller subnetwork of the prior port requires substantially less bandwidth, but potentially misses hosts outside of the subnetwork that could hold informative feature values.
Increasing the seed size increases the fraction of \textit{normalized} services found, since a larger seed size is more likely to encounter uncommon patterns that dominate uncommon ports. However, seed size does {not} substantially affect the fraction of overall services found. We specify in each experiment the seed and scanning step sizes chosen. 

\vspace{3pt}
\noindent
\textbf{Features.}
We specify \system to use 25~features (Table~\ref{table:predictive_features_app}).


\begin{figure*}[t]
  \centering
\begin{subfigure}[]{\columnwidth}
  \centering
	\includegraphics[width=\linewidth]{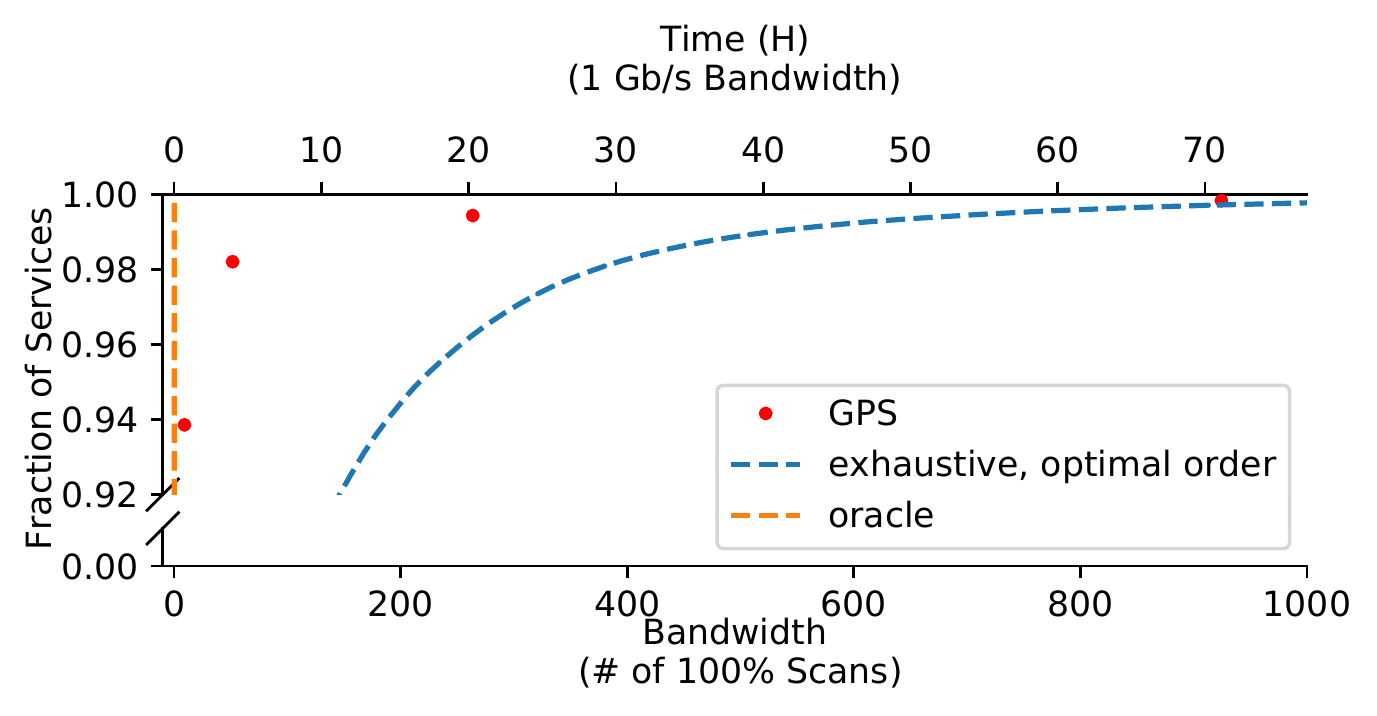}
	\caption{\textbf{Service Discovery (Censys)}: \textnormal{\system finds 94\% of services using $21\times$ less the amount of bandwidth compared to optimal port-order probing (2K ports, 100\% scan, 2\% seed). }}
	\label{fig:seed_gain_fracv}
\end{subfigure}
\hfill
\begin{subfigure}[]{\columnwidth}
  \centering
	\includegraphics[width=\columnwidth]{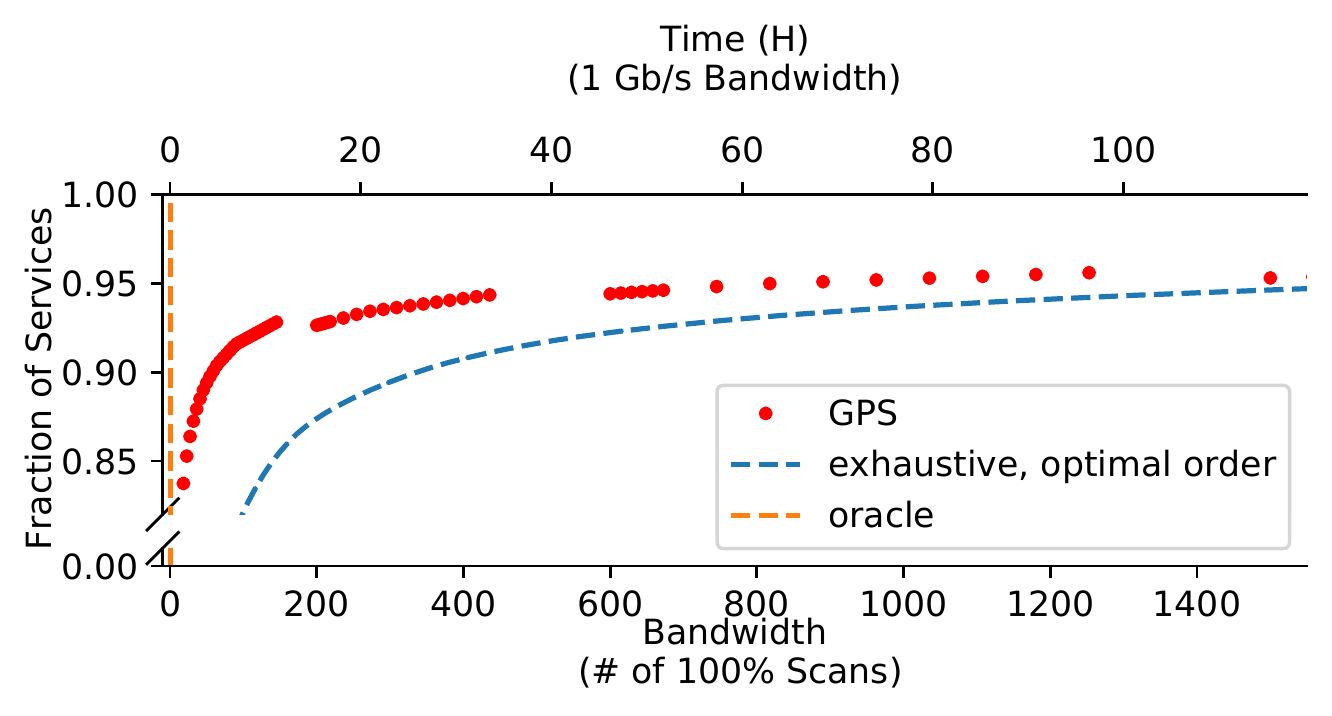}
	\caption{\textbf{Service Discovery (LZR)}: \textnormal{ \system finds 92.5\% and 95\% of all services using $6\times$ less and $2\times$ less bandwidth, respectively, than optimal port-order probing. 
 (all ports, 1\% scan, 0.5\% seed). }}
	\label{fig:seed_gain_fracv_lzr}
\end{subfigure}
\begin{subfigure}[]{\columnwidth}
  \includegraphics[width=\columnwidth]{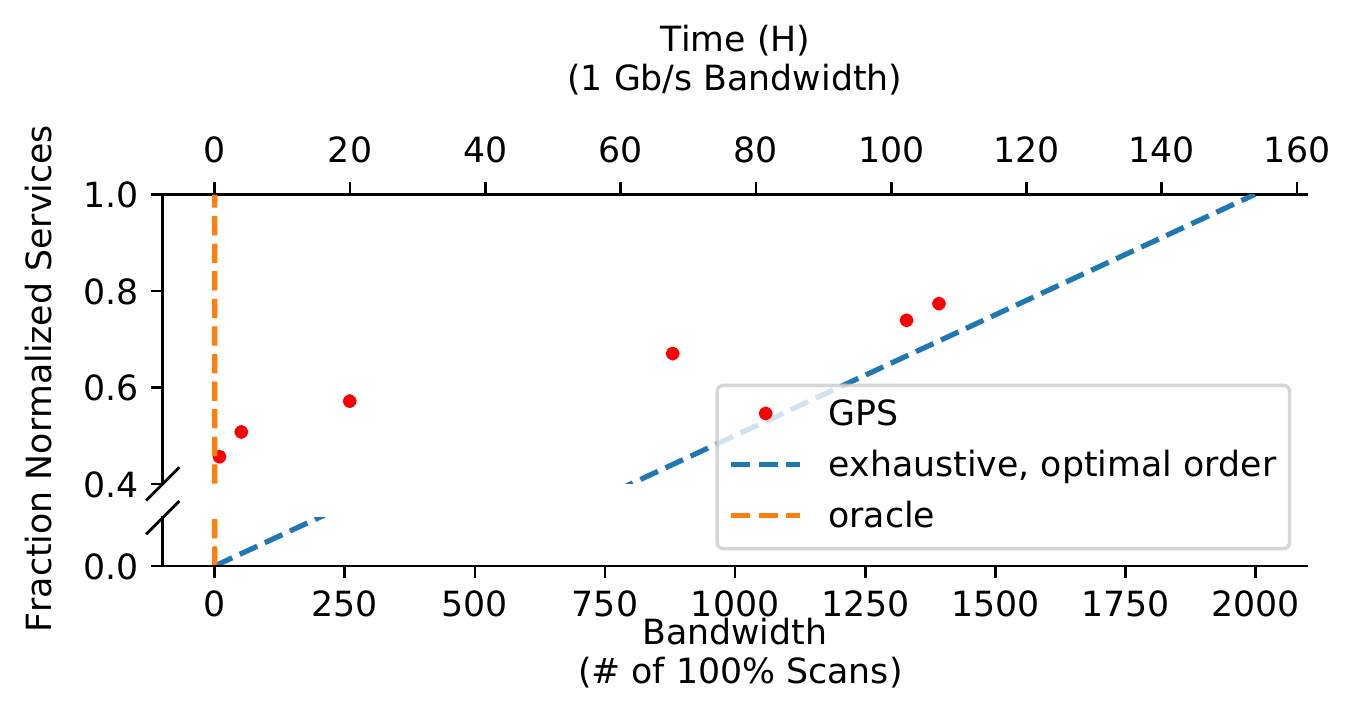}
 \caption{\textbf{Normalized Service Discovery (Censys)}: \textnormal{\system finds 46\% of normalized services across 2K ports using $100\times$ less bandwidth than optimal port-order probing, but when finding 67\% of normalized services, only saves 50\% of bandwidth. (2K ports, 100\% scan, 2\% seed).  }}
  \label{fig:seed_gain}
\end{subfigure}
\hfill
\begin{subfigure}[]{\columnwidth}
  \includegraphics[width=\columnwidth]{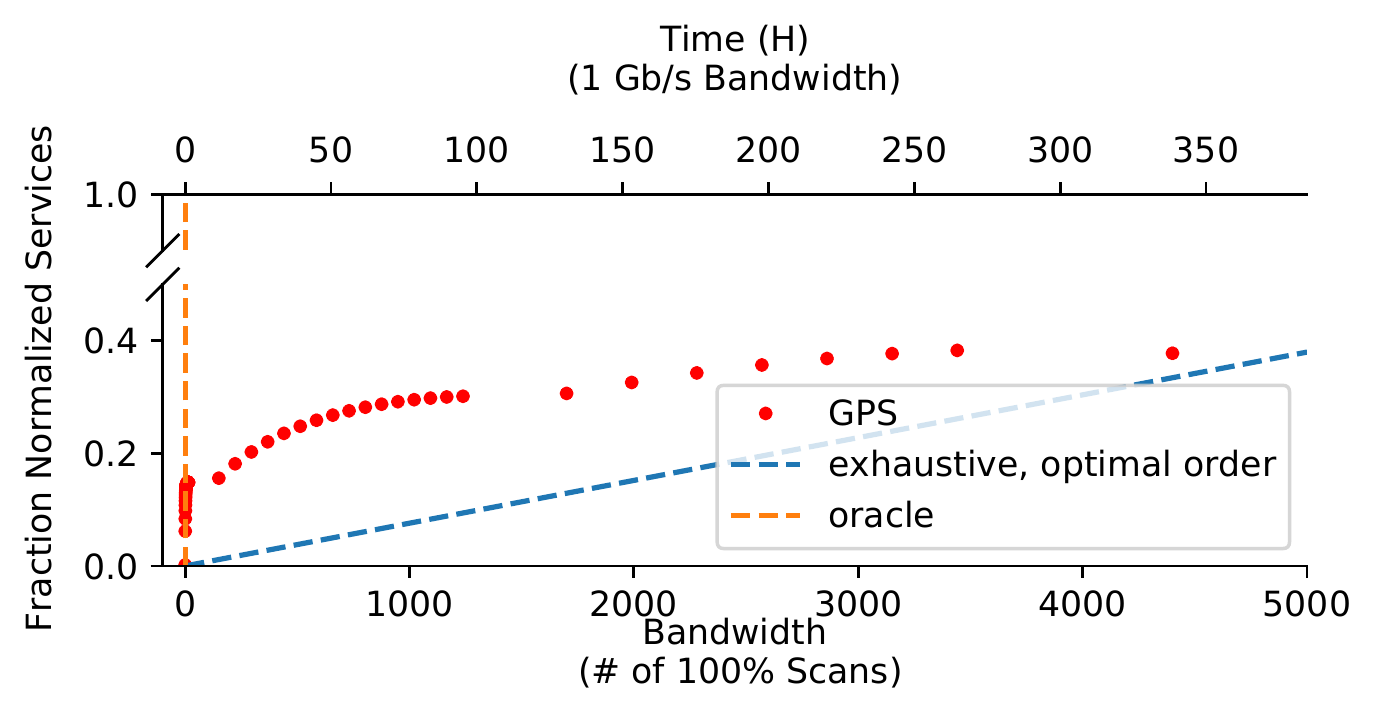}
 \caption{\textbf{Normalized Service Discovery (LZR)}: \textnormal{\system finds 17\% and 38\% of normalized services using $15\times$ and $1.7\times$ less bandwidth, respectively, than optimal port-order probing.  (all ports, 1\% scan, 0.5\% seed).}}
  \label{fig:seed_gain_lzr}
\end{subfigure}
	\caption{\textbf{Finding Services}---%
	\textnormal{Under a variety of training and testing sets, \system is able to find up to 85\% of normalized services and 99.8\% of all services using less bandwidth than optimal port-order probing.
	}}

\label{fig:var_seed_set}
\end{figure*}


\begin{figure}[t] 

  \includegraphics[width=\linewidth]{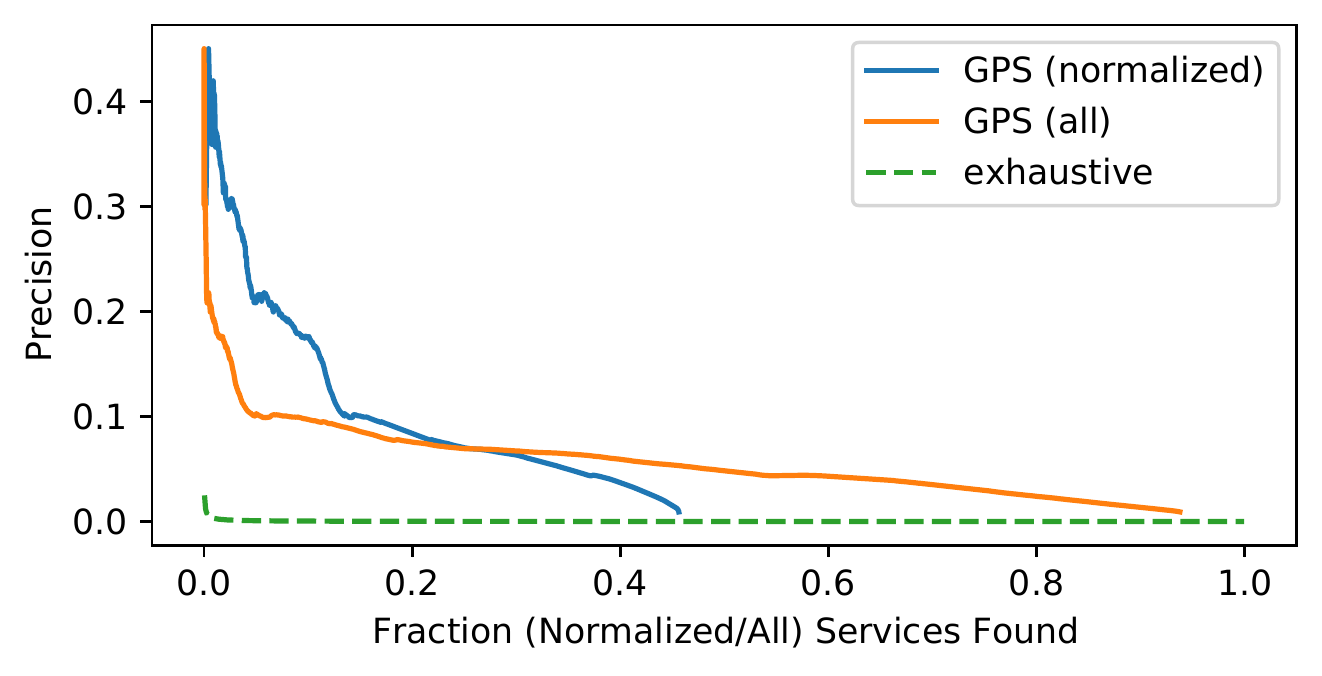}

	\caption{\textbf{\system Precision}---%
     \textnormal{\system finds up to 94\% of all services and 46\% of normalized services while being consistently over an order of magnitude more precise than exhaustive probing.}
}

\label{fig:system_1p_perf}
\end{figure}


\subsection{\system's Coverage Across All Ports}
\label{sub:sec:param}
\label{sub:sub:sec:seed}

\system's objective is to maximize the number of services found across \textit{all} ports. 
To quantify coverage, we use two metrics: fraction of all services found (Equation~\ref{eq:frac_services} in Section~\ref{sub:sec:sys_objective}) and fraction of normalized services found (Equation~\ref{eq:norm_services} in Section~\ref{sub:sec:sys_objective}), which weighs services on popular and unpopular ports equally.
We calculate both metrics across both the Censys (2\% seed set) and LZR (0.5\% seed set) datasets. 
As a reference point, we also plot (1) the bandwidth used by an oracle predictor that knows exactly which services to probe (i.e., 100\% accuracy) and (2) the fraction of services found as a function of bandwidth when exhaustively probing the minimum subset of ports that maximizes service discovery  (i.e., port~80, (80,443), (80,443,7547), etc) . In comparison to exhaustively scanning all ports, this ``optimal port-order'' probing benchmark provides a tighter estimate for the minimum number of ports that must be exhaustively probed in order to find a maximum fraction of services. For example, exhaustively scanning only the 10~most popular ports (i.e., 0.015\% of all ports) is the minimum set of ports that finds 5\% of all services.
We report bandwidth usage in units of 100\% scans (i.e., 3.7~billion packets), in order to easily compare \system performance with exhaustive scanning.

The number of services \system finds depends on the bandwidth budget (Figure~\ref{fig:var_seed_set}).
When evaluating against 100\% IPv4 scans across 2K ports (Censys), \system initially finds 94\% of services using $21\times$ less the amount of bandwidth compared to optimal port-order probing. 
However, since \system scans services in descending order of predictability, predicting services quickly becomes more bandwidth consuming: finding 98.2\% of services across 2K ports only saves $7.6\times$ the bandwidth compared to optimal port-order probing. 
Censys shares that they probe the equivalent of 572~100\% scans every day to curate their dataset; \system uses $3\times10^{11}$ probes, or $6.3\times$ less probes than Censys, to finds 98.2\% of services across 2K ports.   

The bandwidth coverage trade off is exacerbated when finding normalized services: \system finds 46\% of normalized services across 2K ports using $100\times$ less bandwidth than optimal port-order probing but, when finding 67\% of normalized services, saves only 50\% of bandwidth.
When evaluating against 1\% IPv4 scans across all ports (with greater than 2 responsive IP addresses), \system finds 92.5\% and 95\% of all services using $6\times$ less and $2\times$ less bandwidth, respectively, than optimal port-order probing. 
\system's performance drop when evaluating against all ports is likely due to the 4x smaller seed size; \system's performance is nearly identical when using a 0.5\% training seed set across 2K ports (Appendix~\ref{app:seedsize}).

We quantify the fundamental limitations that \system faces when predicting services in Section~\ref{sub:sub:sec:limitations}.





\subsection{\system's Precision Across All Ports}
\label{sub:sub:sec:accEval}
When minimizing Internet-wide scanning's impact on destination networks, we show that \system is over two orders of magnitude more precise than random probing.
\system scans services that are most predictable first (Section~\ref{sub:sec:predicting}).
Thus, \system requires less bandwidth, is more precise, and minimizes time-to-discovery to find the most predictable services at the beginning of the scanning schedule.
We use the Censys dataset, to evaluate against 100\% scans, with a 1\% seed size.
To maximize \system's precision, we configure \system with a small (/20) scanning step size.
We plot in Figure~\ref{fig:system_1p_perf} \system's precision as it continues to find services, and, for comparison, show the precision of  exhaustively probing all ports in the optimal order that prioritizes discovering the most number of services first.

\system is able to find the first 1\% of all services with a 36\% precision---one order of magnitude more precise than exhaustive probing. 
\system finds up to 94\% of all services and 46\% of normalized services while being consistently over an order of magnitude more precise than exhaustive probing. 
For example, \system finds the 94th-percentile of services with $204\times$ more precision than exhaustive probing. 
Note that \system's precision decreases over time as it continues to exhaust its predictions in descending order of predictability. 
Once \system exhausts all predictions, it can be optionally configured to randomly probe the rest of all previously un-probed services, thereby eventually finding 100\% of all normalized services (albeit at a very slow rate). 
\looseness=-1




\subsection{\system vs. Machine Learning}
\label{sub:sub:sec:sysML} 

\begin{figure}[t]
  
\begin{subfigure}[t]{\linewidth}
  \includegraphics[width=\linewidth]{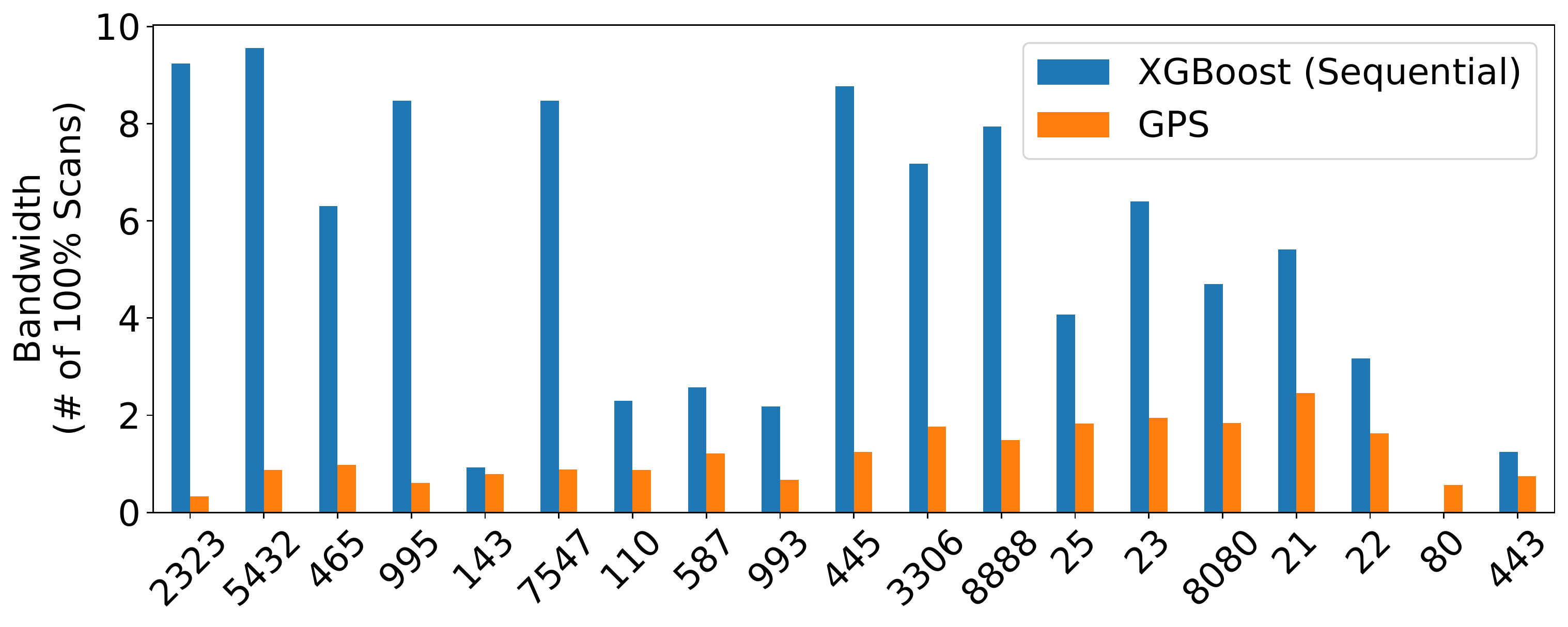}
 \caption{Bandwidth Used to Scan Minimum Set of Predictive Services---\textnormal{\system saves up to 28x more bandwidth than the XGBoost scanner.}}
  \label{fig:sarabi_priors}
  \vspace{3pt}
\end{subfigure}
\begin{subfigure}[t]{\linewidth}
  \centering
	\includegraphics[width=\linewidth]{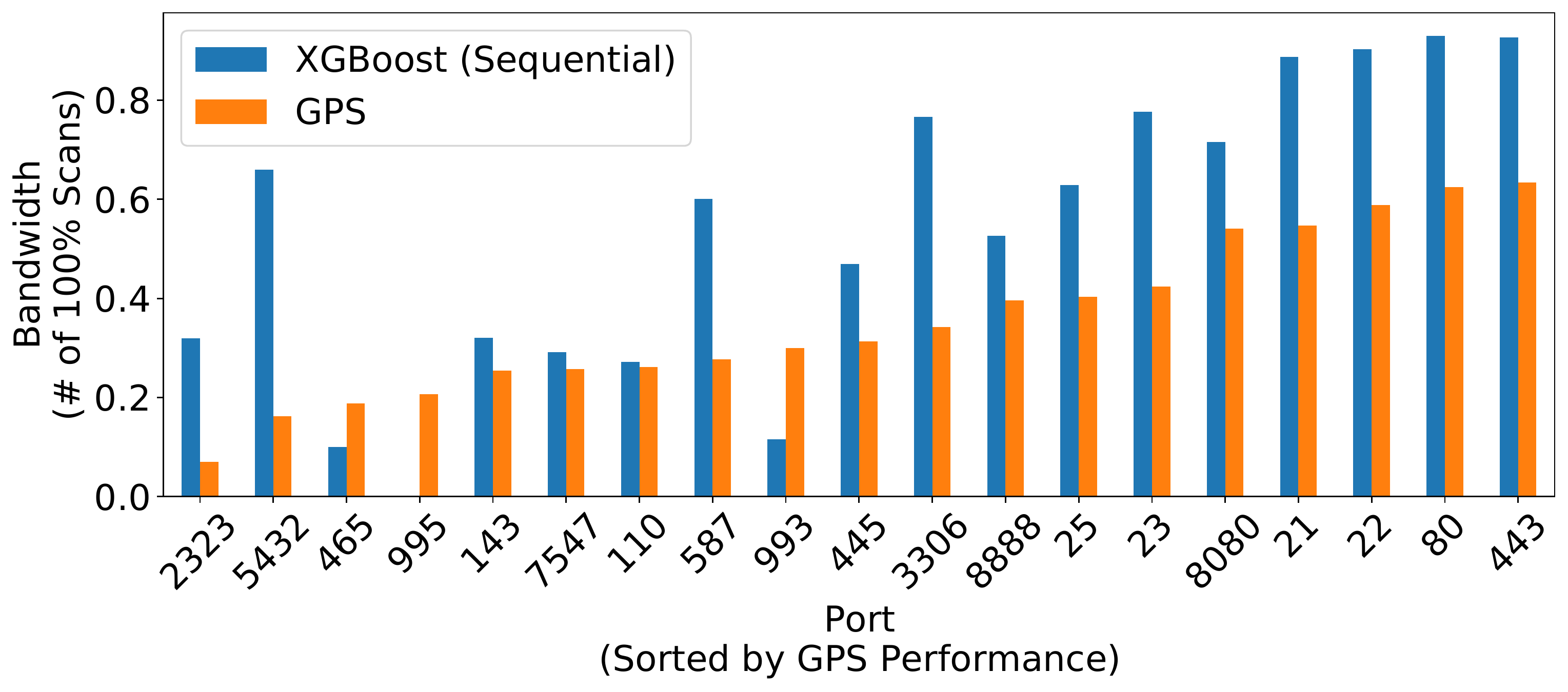}
	\caption{Bandwidth Used to Scan Remaining Services---\textnormal{\system saves more bandwidth than XGBoost scanner when scanning 16 of 19 popular ports. }}
	\label{fig:sarabi_band}
	\vspace{3pt}
\end{subfigure}

\begin{subfigure}[t]{\linewidth}
  \centering
	\includegraphics[width=\linewidth]{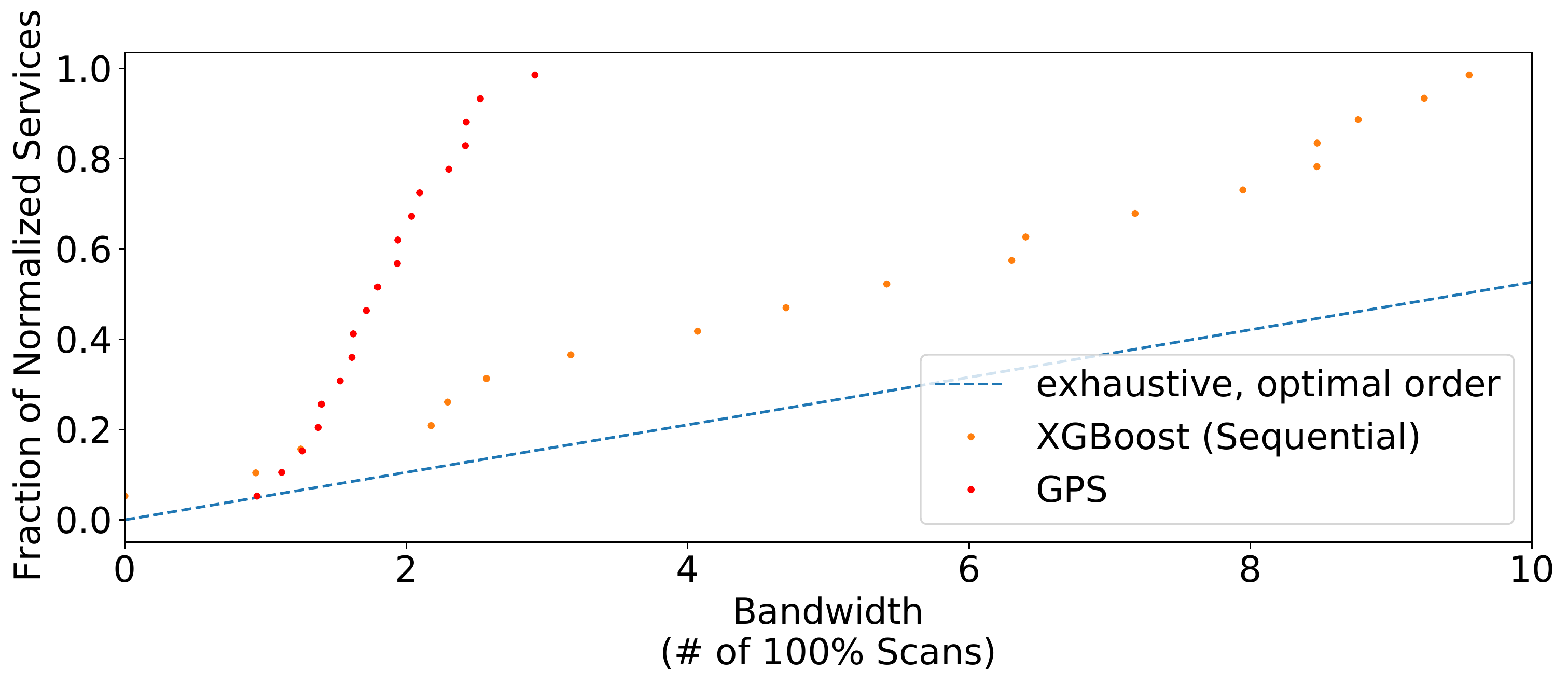}
	\caption{Normalized Service Discovery---\textnormal{\system uses 3x less bandwidth to find 98.5\% of all normalized services than XGBoost scanner.}}
	\label{fig:sarabi_norm}
\end{subfigure}

	\caption{\textbf{\system vs. XGBoost Bandwidth Consumption}%
	}

\label{fig:sarabi_comp}
\end{figure}


As introduced in Section~\ref{sub:sec:prior}, Sarabi~et~al.~\cite{sarabi} use a state-of-the-art XGBoost classifier~\cite{chen2016xgboost} to predict responsive IPs for a given port. 
Unfortunately, the classifier cannot be deployed across all 65K~ports due to (1)~its prohibitively expensive runtime and (2)~the lack of 10~million IP address samples across 99.99\% of all 65K ports to train robust models~\cite{sarabi}. Nevertheless, we benchmark \system against the XGBoost since it is the closest related work. 

\vspace{3pt}
\noindent
\textbf{Methodology.}\quad
Since Sarabi~et~al.'s XGBoost scanner is not open sourced,
 we benchmark \system using the results presented in their paper. 
We use XGBoost scanner's best performing implementation: a sequence of models that follow an optimal ordering of port scanning and use port responses as input features. 
We configure \system to use a seed size of 0.5\% (equivalent to the number of IPs Sarabi et~al.\  use to train their model) from the Censys dataset---also used by Sarabi et~al.\ ---for evaluation. 
We configure \system to use a /16 step size in order to balance coverage and accuracy.
We provide an evaluation for all 19~TCP ports and protocols evaluated by Sarabi~et~al. 
Since their work provides metrics across varying step-sizes of coverage,
we evaluate XGBoost scanner at the maximum coverage level that \system can achieve\footnote{Using \system with a 0.5\% seed size and /16 step size, \system achieves the following maximum coverage: 99.9\% for ports 21,22,80,443, 99\% for ports 23, 25, 119, 445, 465, 587, 993, 3306, 7547, 8080, 8888; 98\% for ports 143, 995, 5432; and 94\% for port 2323. } (i.e., a 98.8\% average coverage). 

\vspace{5pt}
\noindent
Similar to how \system scans a minimum set of predictive services across all hosts to find remaining services, XGBoost scanner also relies on a set of prior scanned IP addresses and ports to predict additional services. 
In Figure~\ref{fig:sarabi_priors} we plot the bandwidth required by the XGBoost scanner to collect all needed prior information (following their optimal scanning sequence) and compare it to the bandwidth required by \system to scan the minimum set of predictive services.
We note that, while Sarabi et~al.\  do not directly reveal their model's prediction accuracy, bandwidth and accuracy are correlated: the less bandwidth needed to accurately predict services, the more accurate the model's predictions are. 

Across all ports, \system requires an average 5.7x less bandwidth to collect its minimum set of predictive services and, at best, 28x less bandwidth when scanning port~2323. 
Scanning port~80 is the only case in which \system requires more bandwidth than the XGBoost scanner; XGBoost scanner does not rely on a minimum set of predictive services to predict services on port~80 (i.e., it only uses network layer features to immediately predict all services). 

Assuming that the minimum set of predictive services has been scanned (e.g., a user originally intended to scan more than one port), we show in Figure~\ref{fig:sarabi_band} the bandwidth required by each system to achieve majority coverage\footnotemark[1] of a target port. 
\system requires less bandwidth than XGBoost when scanning 16 out of the 19 evaluated ports. 
For example, \system requires 4x less bandwidth when predicting hosts on ports~2323 and 5432. 
Ports~995, 993 and 465 are the only ports that XGBoost requires less bandwidth to scan. However, XGBoost's performance gains are not free: the XGBoost scanner require 3--14x more prior bandwidth to collect the minimum set of predictive services for ports~995, 993 and 465 compared to \system (Figure~\ref{fig:sarabi_priors}).
On average, \system requires half the amount of bandwidth to achieve the same coverage of a target port. 
In other words, simple conditional probabilities, on average, achieve a greater accuracy than the XGBoost machine learning classifier when predicting services.

When accounting for all the bandwidth needed to find services, we show in Figure~\ref{fig:sarabi_norm} how \system finds 98.5\% of all normalized services using 3~times less bandwidth than the XGBoost scanner.

\subsection{Computational Complexity}
\label{sub:section:eval_system_implenetation}
\label{sub:sec:comp_compl}

\begin{table*}[t]
\centering
\small
\begin{tabular}{llllll}
\toprule
 & Bandwidth & Computation Time & Wall-clock Time  & Data Processed/Shuffled & Cost  \\
 & & (Single Core) & & & \\
 \midrule
1\% Seed Scan (if needed)  &1.5~Gb/s & -- & 12~Days & -- & -- \\
Seed Scan Upload &20~Mb/s & -- & 3.5~Min & 4~GB & 0\textcent \\
Predicting First Service (PFS) & - & 6~Days 2~Hours & 8~Min (BigQuery) & 4~TB & 13\textcent \\
PFS Download & 1.3~KB/s & -- & 7~Sec & 9.3 KB & 0\textcent \\
PFS Scan & 50 Mb/s & -- & 20~Min & -- & -- \\
PFS Scan Upload & 30 Mb/s & -- &  34~Min & 55~GB & 0\textcent\\
Predicting Remaining Service (PRS)  &--& 3~Days 7~Hours & 5~Min (BigQuery) & 2.5~TB & 62\textcent \\
PRS Download  &18 Mb/s & -- & 8.5~Hours & 547~GB & 0\textcent \\
PRS Scan &50 Mb/s & -- & 8~Hours & -- & -- \\
\midrule

Total Scanning Wall-Time &--  & --&12.3~Days &--&-- \\
Total Download/Upload Wall-Time &-- &-- & 9.1~Hours &--&--  \\
Total Computational Wall-Time& --&--& 13~Min &-- &-- \\

\midrule
Total &-- & 9~Days 9~Hours & 12.7~Days &7~TB & 75\textcent \\

\bottomrule
\end{tabular}
\vspace{3pt}
	\caption{\textbf{\system Performance Breakdown}---%
	\textnormal
	{When using a 1\% seed scan and a /16 scan step size to predict services across all ports, \system bottleneck lies in bandwidth. 
	\system computes all the necessary predictions within 13~minutes when using BigQuery. 
	Due to \system's high scanning precision, both prediction scans use a substantially lower scanning rate (i.e., 50 Mb/s compared to 1.5Gb/s) in order to avoid packet congestion and incoming packet drop.
	}}
	 \label{table:sys_cost}
\end{table*}

Given the appropriate computational resources, \system's parallelizable prediction algorithm allows it to find services across all ports in a constrained amount of wall time.
In this section, we discuss how \system's performance is dependent upon the availability of a seed scan, bandwidth, and computational resources. 
We report the breakdown of \system's performance in Table~\ref{table:sys_cost} when configuring \system to use all 25~features in Table~\ref{table:predictive_features_app}, a 1\% seed scan and a /16 scan step size to predict services across all ports. 

\vspace{3pt}
\noindent
\textbf{Time.}\quad
\system spends time on three categories of tasks: scanning, prediction, and data transfer.
\system's bottleneck lies in bandwidth. Without an existing seed scan, \system requires 12.3~days to perform all scans, which is bounded by existing scanning tools and available bandwidth.
Collecting the initial 1\% seed scan contributes to 97.5\% of all scanning time, due to the low precision when randomly probing (i.e., below 0.00001 for the majority of ports). 
However, if a seed scan is already available, \system can forego collecting the initial seed scan, reducing the overall runtime by 94\%.
In contrast to the seed scan, both prediction scans take only hours to execute, due to their orders of magnitude higher precision than exhaustively probing uncommon ports.
In total, \system predicts 28~billion services that require 8~hours to scan. 

On a single CPU core, \system performs predictions in roughly 9~days and 9~hours---5.6x faster than the XGBoost scanner on a single GPU.
However, \system's wall time can be drastically reduced when marrying its parallelizable prediction algorithm with a highly parallelizable computing environment. 
When implementing \system's prediction algorithm to use BigQuery, \system computes all predictions within 13~minutes of wall-time, \textit{4~orders of magnitude} less time than the XGBoost scanner, while costing only 75~cents. 

Unfortunately, computing over big data on any platform often faces a data transfer bottleneck. Using a serverless platform requires that data be uploaded and downloaded to/from the platform. \system needs a total of 9.1~hours to up/download a total of 606~GB of data to BigQuery. The biggest bottleneck in the up/download process is the up/download bandwidth, which \system experiences to be between 18MB/s--30MB/s when using 24~parallel processes. 
BigQuery does not charge for inbound data transport, leaving the total BigQuery cost to be bounded by computation.
Sarabi et~al.\  do not report the time it takes to load all necessary data into XGBoost scanner's GPU, which is often considered to be the biggest bottleneck for large-scale machine learning computations~\cite{gpuData}.

\vspace{3pt}
\noindent
\textbf{Space.}\quad
The amount of memory \system requires is directly correlated with the size of the seed scan, the number of features used to predict services, and the implementation used to  calculate conditional probabilities.
The seed scan itself---including all of its features---is often quite small; a filtered\footnote{Using the methodology in Appendix~\ref{app:filt_services}.} 1\% IPv4 65K~port seed scan collected by LZR~\cite{lzr} is only 4~GB. 
However, \system requires substantially more memory when predicting services, which is determined by the implementation of conditional probability calculations.
For example, the implementation described in Section~\ref{sub:sec:implementation}---
\JOIN-ing the data on itself to find all pair-wise combinations of an IP's features---increases the memory footprint by at least 50~fold relative to the seed scan and the number of features used.
\system also requires sizeable disk space when outputting the final list of predicted  services; writing the predicted 28~billion IP and port pairs results in 547~GB of output. 

\vspace{3pt}\noindent
Researchers deploying \system will achieve the largest performance gains when using a highly parallelizable computational environment with two orders of magnitude more memory than the initial seed scan size.


\subsection{Which features are most predictive?}
\label{sub:sec:predictive_features}

\begin{table}[t]
\centering
\small
\begin{tabular}{lll}
\toprule
Feature & Normalized & Services \\
& Services & \\\midrule
( Port, Port\textsubscript{Protocol} )   &	18.7\% & 2.0\% \\
 Port   &	14.1\% & 2.0\% \\
 ( Port, Port\textsubscript{HTTP\: Header} )  &	9.7\% & 2.0\% \\
( Port, Port\textsubscript{ASN}, Port\textsubscript{HTTP-Body-Hash} )   &	7.7\% & 2.0\% \\
( Port, Port\textsubscript{HTTP-Body-Hash} )   &	6.1\% & 2.0\% \\
\bottomrule
\end{tabular}
\vspace{3pt}
	\caption{\textbf{Top~5 Predictive Features}---%
	\textnormal{A port's protocol is the most predictive feature, predicting 18.7\% of normalized services.
	}
	 }
	 \label{table:predictive_features}
\end{table}

By using conditional probabilities, it is simple to understand what features are most predictive of Internet services. 
When running \system using a 1\% seed set to predict services across all ports in the Censys dataset, \system selects 402K unique feature \textit{values} as being most predictive of a service. 
Information found when using the HTTP protocol (e.g.,  HTTP header, HTTP Body hash) is most predictive compared to every other protocol, contributing to 45\% of the most predictive features values.
We present the top-5 most predictive feature candidates that \system identifies in Table~\ref{table:predictive_features}.
Across 18.7\% of normalized services, the protocol (e.g., SSH) running on a host's port (i.e., (Port, Port\textsubscript{Protocol})) is  most predictive of another port being responsive. 

\system identifies 64~unique tuples of feature \textit{values} that are most predictive of services, which include the interaction of application-layer and network-layer features such as: (ASN, TLS certificate) (4.4\% normalized services), (ASN, SSH Key) (1.2\%), VNC name (0.4\%), and (ASN, FTP banner) (0.24\%). 
For example, 95\% of hosts in Distributel Network (ASN\,1181) that respond on port 23 with the telnet banner ``Telnet service is disabled or Your telnet session has expired due to inactivity...'' host HTTP content on port~8082; and 98\% of hosts in Bizland (ASN\,29873), which respond with an IMAP banner requesting TLS on port~143, also host SSH on port~2222. 
These findings show that \system's ability to model the interaction of features (Section~\ref{sub:sub:sec: identify_important}), and decision to not exclude any feature candidates, helps predict services  on various networks and ports. 






\section{Limitations}
\label{sub:sub:sec:limitations}
While \system dramatically shifts the barrier for scanning all Internet services, there remain existing challenges.  

\vspace{3pt}
\noindent
\textbf{Pattern Mining.}\quad
\system is bounded by the features it is configured to use (Appendix~\ref{app:features}) and the resulting patterns it mines. If, for example, detecting a pattern relies on collecting features from the union of multiple port responses, \system may not detect the pattern due to the computationally expensive nature of calculating correlations between more than two ports. Nonetheless, with the advent of increased availability of computational parallelism, \system's algorithm is modular and easily extensible to add additional feature correlation computations. 

\vspace{3pt}
\noindent
\textbf{IPv6.}\quad
\system relies on exhaustive scans to obtain a set of responsive IP addresses to use for service predictions: this approach does not work for IPv6 due to the larger search space. However, given known IPv6 addresses that respond on at least one port, \system can be used to predict other responsive services on the known IPv6 addresses. 

\vspace{3pt}
\noindent
\textbf{Random Host Configuration.}\quad
While services often exhibit predictable patterns, random configurations of hosts will always present a limitation for predicting services.
For example, the manual of the most common (21\%) IoT device found in our LZR scan, FRITZ!Box, 
states that ``for security reasons, FRITZ!Box sets up a random TCP port for HTTPS when internet access via HTTPS is enabled''~\cite{fritzSecurity, fritzManual}. 
Furthermore, routers can easily port-forward services through random ports~\cite{portForward,routerRandom}.
We find in our LZR scan that at least 55\% of services are likely being port-forwarded (i.e., different TTL values returned across all services being hosted) across 99\% of the most uncommon ports. 

To understand how random host configurations impact \system, we set up an experiment in which we: (1) Use a 95\% seed set to predict the remaining 5\% of services of the Censys dataset, thereby assuming that nearly all patterns are known beforehand; (2) Count all services on an IP as being discovered the moment at least one service has been discovered on an IP,  thereby assuming that feature correlations are 100\% available and 100\% accurate; and (3) Specify the largest scanning step size (/0) to maximize the total fraction of normalized services found.
Under these ideal conditions,  80\% of all normalized services can be discovered using less bandwidth than exhaustive scanning, a percentage slightly lower than \system due to the small size of the random test set.
These results illustrate how, in a ``real-world'' setting, \system performs near what is best achievable and illustrates what fundamental limitations lie ahead for any intelligent Internet-wide scanning system. 
\section{Conclusion}

In this work, we introduced \system, an intelligent scanning system that scalably and efficiently predicts services across all IPs and ports with no prior knowledge. 
We demonstrated how a seemingly simple predictive framework based conditional probabilities can perform orders of magnitude faster and with more accuracy than the leading machine learning implementation.
\system finds 92.5\% of services across all ports using $131\times$ less bandwidth than exhaustive scanning, while being $204\times$ more precise.
By releasing \system as an open source tool, we hope that the research community will now be able to find the billions of previously-missed services, at a fraction of the cost of exhaustive scanning.
\section*{Acknowledgements}
The authors thank Tatyana Izhikevich, Katherine Izhikevich, Kimberly Ruth,
Deepak Kumar, Pratiksha Thaker, Deepti Raghavan, members of the Stanford
University security and networking groups, our shepherd, Lixia Zhang, and the anonymous
reviewers for providing insightful discussion and comments.
This work was supported in part by the National Science Foundation under award
CNS-1823192, Google., Inc., the NSF Graduate Fellowship
DGE-1656518 and a Stanford Graduate Fellowship.

{\footnotesize  \bibliographystyle{abbrv}
\bibliography{reference}}
\appendix

\setcounter{secnumdepth}{0}
\section{Appendix}


Appendices are supporting material that has not been peer-reviewed.

\setcounter{secnumdepth}{1}
\section{Recommendation Systems for Intelligent Scanning}
\label{app:rec_sys}
Proprietary recommendation systems have successfully recommended millions of items to millions of users at Netflix~\cite{gomez2015netflix}, Spotify~\cite{mcinerney2018explore}, and Amazon~\cite{ma2020temporal}. 
The model's success inspires us to use a recommendation system to recommend/predict responsive ports for IP addresses. 
We explore over 30~different open source recommendation models~\cite{recMic} and find that while existing recommendation systems allow features to be assigned to users (i.e., IPs) and/or items (i.e., ports), they do not directly allow for features to be assigned to their interaction (i.e.,  IP,Port tuples). 
Consequently, application layer features for every service cannot be easily represented.
Further, the majority of models do not support adding new features, which may have been discovered as a result of a successful prediction, without the undesirable computational overhead of re-training the model.

To evaluate if open source models can successfully predict services, we extend a popular open-sourced hybrid recommender system, LightFM~\cite{kula2015metadata} to predict a responsive port, given an IP address.
We provide the implementation of the model, including all of its parameters, on GitHub\footnote{\url{https://github.com/stanford-esrg/recommender-system-gps}}.
Given the framework's inability to assign features to specific services (i.e., (IP, Port) pairs), we are only able to encode features for IP addresses and ports. 
We experiment with assigning different network layer features (e.g., autonomous system, /16 subnetwork, /20 subnetwork), to every IP address and assigning a binary feature (designating whether the port number is IANA assigned) to every port. 
We train the model on the LZR dataset across all 65K ports (using an 0.8\% IPv4 seed set) and have it generate 100 port predictions for every IP address in the test set (i.e., similar to generating 100 ``100\% prediction scans''). 
The model finds a maximum of 47\% of all services---consistently performing worse compared to exhaustively probing ports in an order the prioritizes finding the most number of services---and 1.5\% of normalized services. 
Due to the feature constraints that recommendation systems impose and this model's poor performance, we choose to stop pursuing adapting recommendation systems for predicting services across all ports.

\setcounter{secnumdepth}{2}
\section{Filtering For Real Services}
\label{app:filt_services}

Prior work has shown that a substantial number of IP addresses host ``pseudo services''~\cite{censysServiceSinks}.
Pseudo services are often HTTP or HTTPS webpages that have successfully loaded, but display a message stating that no services exists on the webpage itself.  
We conduct a LZR~\cite{lzr} scan across all 65K ports on a random 1\% subset of the IPv4 address space in March~2021 and find that across 96\% of all 65K ports, the vast majority of services belong to a host that serve pseudo services on greater than 1,000 contiguous ports.

To ensure \system does not learn to predict ``pseudo services,'' \system's seed set must filter all pseudo services.
Over 80\% of pseudo services are simply filtered by first, removing expected dynamic fields (e.g., HTTP date field, HTTP Cookie field, TLS random bytes) in the data and secondly, removing all services on the host that share the same filtered data.
However, the long tail of pseudo services are much harder to fingerprint and filter. 
For example, data containing an unexpected random string (e.g., ``Incident ID'', timestamp), results in the data and content-length to slightly differ.
Existing filtering solutions are often complex and incomplete~\cite{ma2006unexpected}.
Thus, we filter any host that serves more than 10 services---a method identifies pseudo services with 100\% recall and 99\% precision.
All results in this work assume that pseudo services are filtered.

\setcounter{secnumdepth}{3}
\section{Network-Layer Feature Candidates}
\label{app:features}

The final implementation of \system is configured to use only the network-layer features that are most predictive of the majority of services.
Reducing the number of network-layer features reduces the computational complexity of \system's predictive algorithm. 
To determine which network-layer features are most predictive of service presence, we initially configure \system to use all subnetwork sizes between /23--/16, and the autonomous system number.
\system builds a probabilistic model using a 0.5\% seed set from the LZR dataset, which we use to analyze which network features are most predictive across all services. 
We show in Table~\ref{table:predictive_features_net} how the IP's ASN and /16 subnetwork are most predictive of service presence. 
It is no surprise that larger subnetworks are more predictive of service presence: larger subnetworks are more likely to be shared amongst multiple hosts in the seed set.

\begin{table}[t]
\centering
\small
\begin{tabular}{ll}
\toprule
Network Feature & \% Services \\ 
 & Most Predictive  \\ 
\midrule
ASN & 36\% \\
/16 & 20\% \\
/18 & 8\%\\
/19 & 8\% \\
/17 & 8\%\\
/20 & 7\%\\
/21 & 6\%\\
/22 & 4\%\\
/23 & 3\%\\
\bottomrule
\end{tabular}
\vspace{3pt}
	\caption{\textbf{Network Features}---%
\textnormal{When configuring \system to use the /16--/23 subnetworks of an IP address, the ASN and /16 of an IP address are most predictive for the majority of services. }
	 }
	 \label{table:predictive_features_net}
\end{table}

\setcounter{secnumdepth}{3}
\section{Parameter Tuning}
\label{app:parameter_tune}

\subsection{Varying Step Size}
\label{app:stepsize}

\begin{figure}[t] 

  \includegraphics[width=\linewidth]{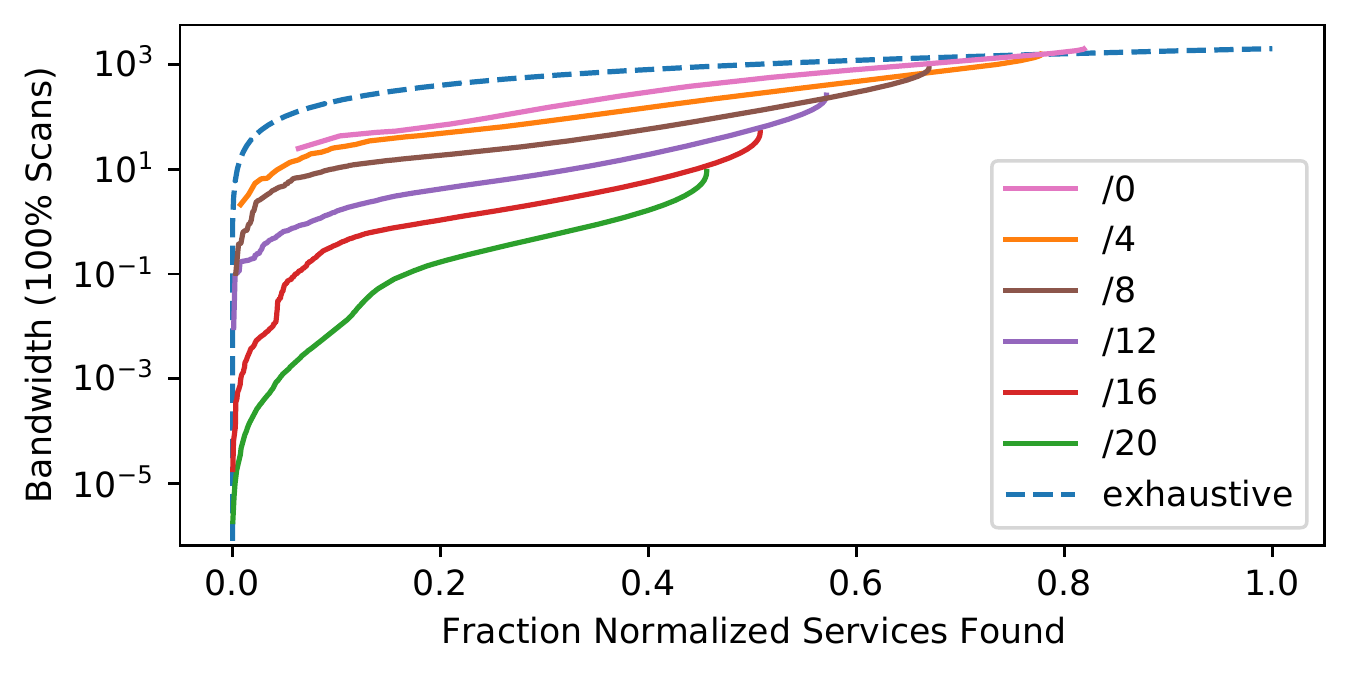}

	\caption{\textbf{Varying Step Size (Censys)}---%
	\textnormal{A smaller scanning step saves more bandwidth when initially finding services, but ultimately finds less services compared to a larger scanning step size. }
	}

\label{fig:system_1p_perf_app}
\end{figure}
\system minimizes bandwidth during prediction by regulating the scanning step size. 
As the step size decreases, the precision of finding services increases, causing the overall bandwidth required to decrease. 
For example, as seen in Figure~\ref{fig:system_1p_perf_app},  finding the first 25\% of normalized services using a scanning step size of /12 uses one order of magnitude more bandwidth compared to a scanning step size of /20. 
No \system configuration finds more than 82\% of normalized services with a bandwidth usage better than exhaustive probing.

\subsection{Varying Seed Size}
\label{app:seedsize}

\begin{figure}[t]
  
\begin{subfigure}[t]{\linewidth}
  \includegraphics[width=\linewidth]{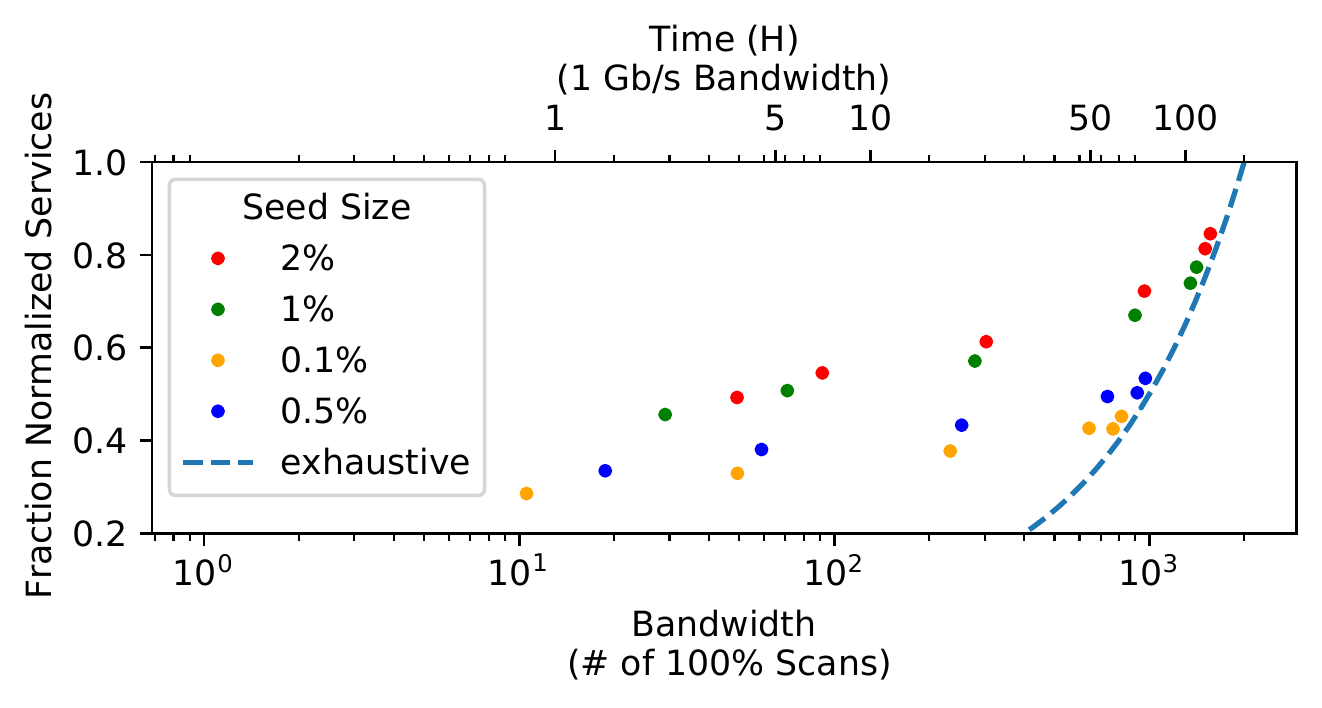}
 \caption{Normalized Service Discovery}
  \label{fig:seed_gain_app}
\end{subfigure}

\begin{subfigure}[t]{\linewidth}
  \centering
	\includegraphics[width=\linewidth]{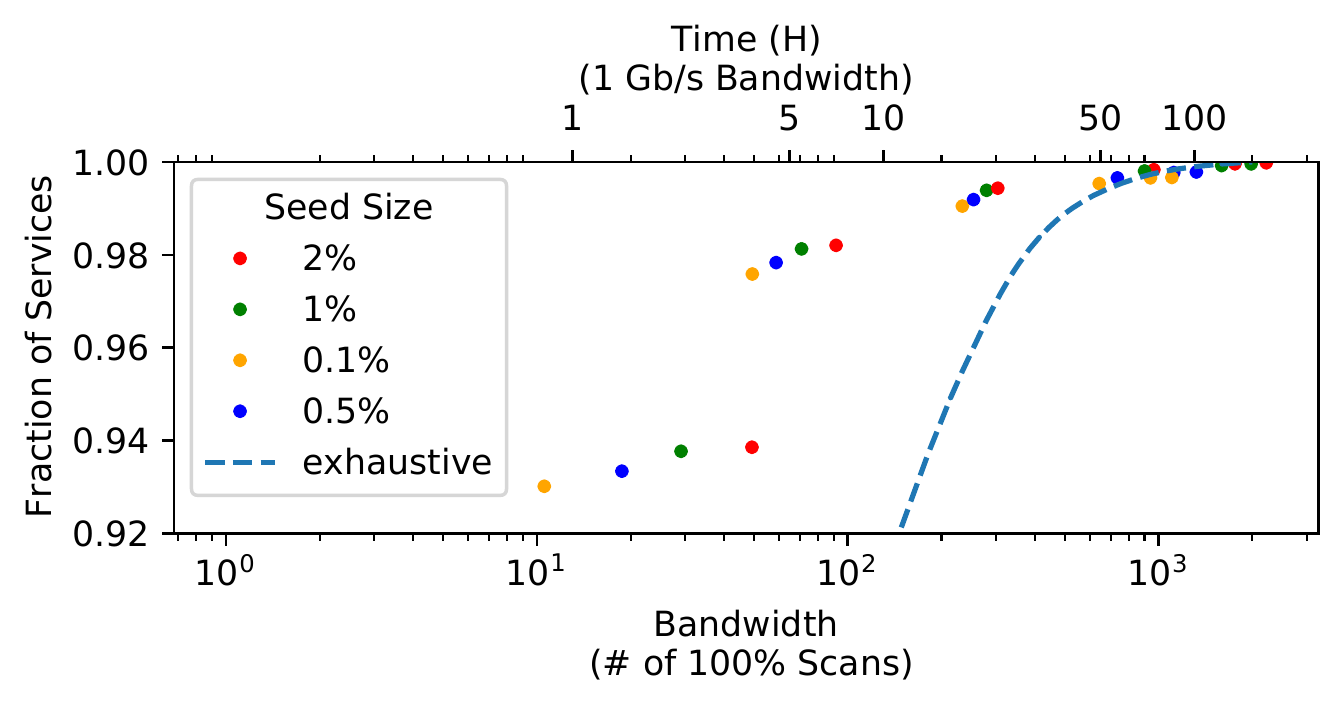}
	\caption{Service Discovery}
	\label{fig:seed_gain_fracv_app}
\end{subfigure}

	\caption{\textbf{Varying Seed Size (Censys)}---%
	\textnormal{A larger seed size increases the fraction of normalized services \system finds, but does not substantially impact the fraction of all services \system finds. 
	}}

\label{fig:var_seed_set_app}
\end{figure}

The seed size regulates the amount of unique service patterns \system has already seen. 
We plot the amount of bandwidth needed (including collecting the seed size) and fraction of services found in Figure~\ref{fig:var_seed_set_app}.
When optimizing to find normalized services, Figure~\ref{fig:seed_gain_app} shows that, for a bandwidth budget above 30~100\% IPv4 scans, a 2\% IPv4 seed size always finds the most fraction of normalized services compared to smaller seed size.
This indicates that patterns found in a larger seed scan are crucial for finding normalized services. 
However, when optimizing to find the largest fraction of services, smaller seed sizes are sufficient (Figure~\ref{fig:seed_gain_fracv}), indicating that the most predictive patterns for finding popular services can be detected with a small seed size.

\end{document}